\documentclass{article}
\usepackage{appendix}
\usepackage{authblk}
\usepackage{xspace}
\usepackage[compat=1.1.0]{tikz-feynman}
\usepackage[margin=1.0in]{geometry}
\usepackage{caption}
\usepackage{subcaption}
\usepackage[normalem]{ulem}
\usepackage{hyperref}
\usepackage{amssymb}

\newcommand{\gf}{\textsc{Geant}4\xspace}
\newcommand{\mg}{\textsc{MadGraph/MadEvent}\xspace}
\newcommand{\aprime}{\ensuremath{A^\prime}\xspace}

\usepackage{acronym}
\acrodef{ww}[WW]{Weizs\"{a}cker-Williams}
\acrodef{dm}[DM]{Dark Matter}
\acrodef{db}[DB]{dark bremsstrahlung}
\acrodef{dmg4}[DMG4]{\cite{dmg4}}
\acrodef{g4db}[G4DarkBreM]{\gf Dark Bremmstrahlung from \mg}
\acrodef{sm}[SM]{Standard Model}
\acrodef{mg}[MG/ME]{\mg}
\usepackage{url}

\title{Simulation of Dark Bremsstrahlung in \gf}

\author[a]{Tom Eichlersmith}
\author[a]{Jeremiah Mans}
\author[b]{Omar Moreno}
\author[a]{Joseph Muse}
\author[a]{Michael Revering}
\author[b]{Natalia Toro}

\affil[a]{University of Minnesota Twin Cities, Minneapolis, MN 55455, USA}
\affil[b]{SLAC National Accelerator Laboratory, Menlo Park, CA 94025, USA}

\begin{document}

\maketitle

\begin{abstract}
    A technique for the simulation of dark bremsstrahlung for electrons and muons in \textsc{Geant}4 is presented. The total cross section is calculated using the Weizs\"{a}cker-Williams approximation and the outgoing kinematics are produced by scaling events produced in \textsc{MadGraph/MadEvent} to lower incident lepton energies, allowing the simulation to account for thick targets and lepton sources without fixed energies.
    Compared with dedicated samples produced at an arbitrary particle energy, typical precision of better than 5\% is achieved.
\end{abstract}

\section{Introduction}
\label{sec:introduction}

The last decade has seen a growth of interest in the idea of a \emph{dark sector} neutral under \ac{sm} forces at the MeV to GeV mass scale \cite{Alexander:2016aln, Lanfranchi:2020crw, Agrawal:2021dbo, Battaglieri:2017aum, BRN, SnowmassRF6}.  Stable matter in a dark sector is an attractive dark matter candidate \cite{Essig:2011nj,Izaguirre:2015yja} and dark sectors can also play a role in resolving other outstanding problems in particle physics as reviewed e.g.~in \cite{Lanfranchi:2020crw,SnowmassRF6}.  Accelerator-based production of dark sector particles is an essential avenue to probing this landscape, highlighted in the 2018 Dark Matter New Initiatives Basic Research Needs Report \cite{BRN} as both a robust probe of sub-GeV thermal dark matter \cite{Izaguirre:2015yja,RF6_BI1} and a unique window on unstable particles in the dark-sector and the interactions they mediate.   General symmetry arguments imply that the leading interactions of \ac{sm} and dark-sector particles should be through a small set of \emph{portal} interactions \cite{Alexander:2016aln,Lanfranchi:2020crw,SnowmassRF6}. Among these, the kinetic mixing of a spin-1 \emph{dark photon} with the \ac{sm} photon \cite{Holdom:1985ag, Okun:1982xi, Galison:1983pa}, which induces a weak dark-photon coupling to ordinary electromagnetic charges, is the most extensively studied due to the model's simplicity, compatibility with simple light dark matter models, and experimental accessibility. 

Electron and muon beam fixed-target experiments are among the first proposals to search for dark photons decaying back to \ac{sm} matter \cite{Fayet:2007ua, Reece:2009un, darkgauge}, and have also emerged as especially powerful probes of dark matter production.  Therefore, accurate modeling of dark photon production in these experiments is an important problem. While several simulation packages have been developed for dark photon production from proton beams \cite{Buonocore:2018xjk,deNiverville:2016rqh}, the options for electron and muon beams are limited. 

Several reactions for dark photon production in lepton beams have been considered in the literature.
Over most of the dark photon mass range being explored, the most relevant process is \ac{db}: emission of a dark photon off a beam-originated lepton as it recoils off the electromagnetic field of a nucleus in the target. \ac{db} tends to dominate over other reactions except in specific mass ranges. For example, annihilation of beam-induced positrons off atomic electrons \cite{Marsicano:2018glj,Marsicano:2018krp} is significant when the dark photon mass is near the positron-electron center-of-momentum energy. For missing energy experiments, the impact of this process is typically muted by the experimental requirement for significant missing energy combined with the necessary production mechanisms for positrons which leave a substantial energy visible in the balance of the electromagnetic shower.  For electron beams, photoproduction of vector mesons that mix with the dark photon becomes prominent for dark photon masses $\geq0.5$ GeV \cite{Schuster:2021mlr}.  Given the complex total cross-section dependence of the meson processes on beam-electron energy, future work would be required to accurately include these effects in a similar simulation tool.

\ac{db} of an on-shell dark photon is a $2\rightarrow 3$ process, with a dark photon, recoil lepton, and recoiling nucleus in the final state.  Although the recoiling target nucleus is generally quite non-relativistic, momentum transfer to the nucleus upsets the transverse momentum balance between the recoil lepton and dark photon, and must be included in a full calculation.  Experiments in lepton beams make use of the detailed kinematics of the dark photon and/or recoil lepton for both their design and signal detection.
To correctly model these experiments,  one must accurately model \emph{both} the kinematics of the \ac{db} process itself and electron interactions before and after the \ac{db} occurs. This need is especially acute in missing momentum experiments with moderately thick targets, such as the full run of the proposed LDMX experiment \cite{LDMX:2018cma,LDMX:2019gvz,Akesson:2022vza}, which are sensitive to the energy and angle of the recoiling electron as well as to the contributions of ordinary photon bremsstrahlung to the final-state energy. 
Most approaches to \ac{db} simulation in the literature sacrifice one or the other aspect of modeling.

Collider-motivated tools like \ac{mg}, with the addition of nuclear form factors, compute fully differential leading-order cross-sections for the \ac{db} accurately; however, {\sc MadEvent} samples the squared amplitude produced by {\sc MadGraph} on the basis of defined input kinematics for a charged particle. Typically, these \ac{db} events are then input into a \gf simulation as the initial event vertex by placing them within the detector's target volume. This neglects any interactions of the beam particles upstream of the target volume and oversimplifies how the beam could interact with the target itself, especially when the target is a substantial fraction of a radiation length or larger.
Another approach is to embed a highly simplified, \ac{ww}-based calculation of the final-state kinematics into \gf \cite{dmg4}, which accurately models the electromagnetic interactions of an incident lepton prior to \ac{db}, but suffers from significant approximation errors in the modeling of the \ac{db} itself (most importantly, this implementation always takes the final-state electron and dark photon to be produced forward, where in fact they typically carry sizable transverse momenta \cite{Izaguirre:2014bca}). This work presents an alternative approach which avoids either of these oversimplifications by embedding in \gf a model of the \ac{db} process that uses the \ac{ww} approximation for the total cross section \emph{and} models final-state kinematics accurately by scaling the energy and transverse momentum using \ac{mg} events. The package introduced in this manuscript will be referred to as \acs{g4db}.

The simulation of the \ac{db} process in the interaction of electrons or muons with target nuclei has been implemented in the \ac{mg} package \cite{madgraph_2014, darkgauge}. The \ac{db} process proceeds with a generic dark sector mediator given as \aprime, as shown in Fig.~\ref{fig:aprimefeynman}. For simulation purposes, it is assumed \aprime is non-interacting after generation. \ac{mg} provides an accurate implementation of the underlying matrix element based on an input field theory model. The model used here includes the \aprime coupling to electrons (the coupling to individual nucleons in the target is not included) and assumes the kinematics of coherent nuclear scattering, which dominates for most dark photon masses. The nucleus coupling to the \aprime is assumed to have a momentum-dependent form factor that includes both coherent and incoherent components, based on \cite{darkgauge,Kim:1972gw}. The code used in \ac{mg} cannot be simply integrated into a detector simulation due to the difficulty of implementing the phase-space integrals and the event processing time.

Simulations of the \ac{db} process within a detector model require calculation of the total cross section for a range of energies and materials to determine the probability of interaction as well as the differential cross section as a function of outgoing lepton kinematics in order to accurately simulate the deflection caused when the interaction occurs. While constant offsets in the total cross section can be removed by simple event re-weighting, energy-dependent variations would require complete information of the lepton energy at each point in its trajectory before the \ac{db} occurs to calculate the correct re-weighting factor. Comparing the results of the \ac{ww} approximation with the results from the \ac{mg} calculation, the approximation has differences of $\mathcal{O}(10\%)$ in the overall scale but provides a reasonable estimate for the energy dependence of the total interaction cross-section, as shown in Fig.~\ref{fig:xsec_comp} for electrons with energies $\mathcal{O}(50)~$GeV interacting with tungsten and muons with energies $\mathcal{O}(500)~$GeV interacting with copper. However, as noted by previous researchers \cite{Liu:2016mqv}, the approximation can be quite inaccurate in predicting the differential cross-section as a function of outgoing \aprime or lepton kinematics. As mentioned previously, we instead present a strategy to scale \ac{mg} events from a library directly into \gf for simulation purposes.

\begin{figure}[!htb]
\centering
\begin{tikzpicture}
  \begin{feynman}
    \vertex (a) {$\ell$};
        \vertex [right= of a](d);
    \vertex [below=of d, blob,label={below:$N$}] (e) {};
    \vertex [above right= of d] (f);
  \vertex [right= of d] (b);
  \vertex [above right= of b] (g) {$\ell$};
  \vertex [below right= of b] (h) {\aprime};
    \diagram*[large] {
      (a) -- [fermion] (d),
      (d) -- [boson] (e),
      (d) -- [fermion] (b),
      (b) -- [fermion] (g),
      (b) -- [boson] (h),
    };
  \end{feynman}
\end{tikzpicture}
    \caption{Feynman diagram illustrating a \ac{db} process. A light lepton ($\ell=e,\mu$) interacts with a target nucleus (N) by exchanging a SM photon followed by the emission of a generic dark sector mediator (\aprime), which is assumed non-interacting after production.}
    \label{fig:aprimefeynman}
\end{figure}
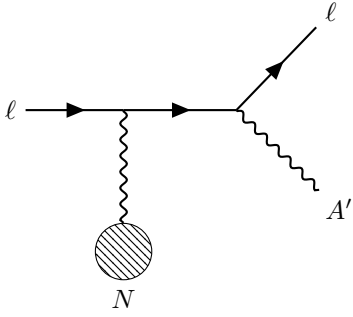
\begin{figure}[!htb]
    \centering
    \begin{subfigure}[b]{0.45\textwidth}
        \centering
        \includegraphics[width=\textwidth]{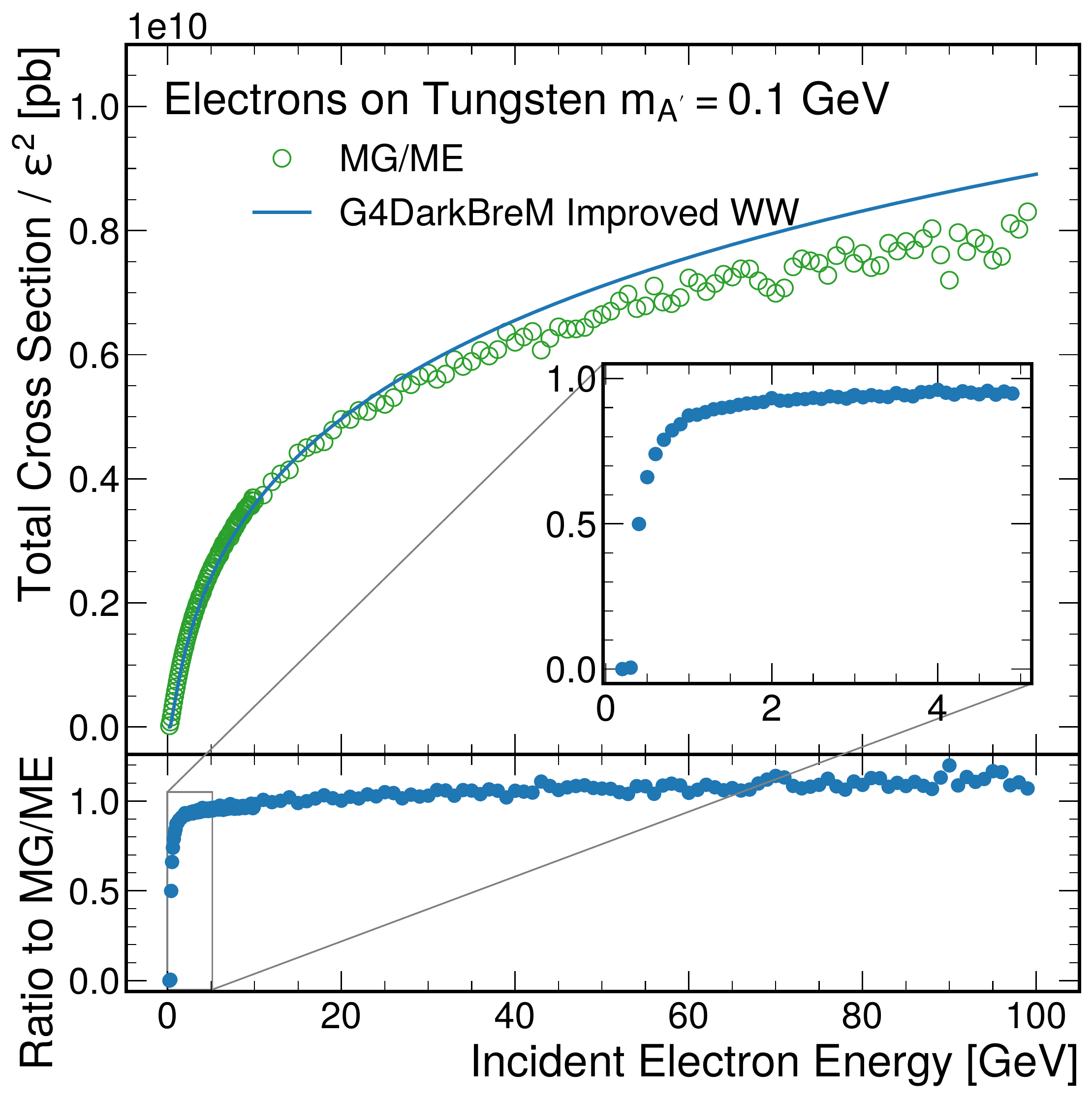}
    \end{subfigure}
    \begin{subfigure}[b]{0.45\textwidth}
        \centering
        \includegraphics[width=\textwidth]{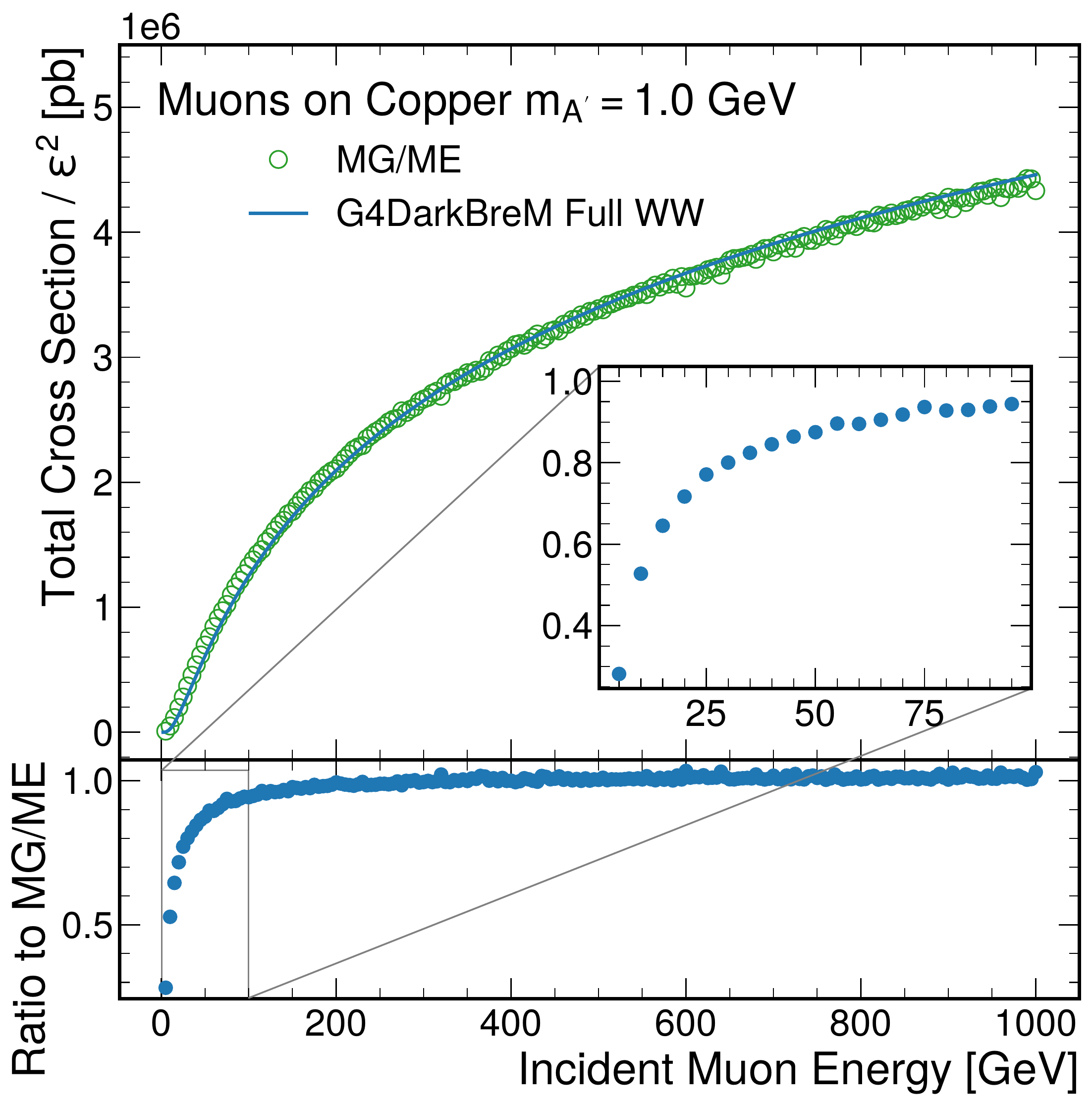}
    \end{subfigure}
    \caption{Comparison between the \ac{ww} approximation as calculated by the \acs{g4db} package (this work) and the total \ac{db} cross section calculated by \ac{mg} for electrons incident on tungsten with the mass of \aprime ($m_{\aprime}$) taken as $0.1$~GeV and muons incident on copper with $m_{\aprime}$ taken as $0.2$~GeV. Other variations of the \ac{ww} approximation are available in this package and detailed in Appendix \ref{sec:docs:xsec}. For the simulation process, the critical characteristic for the total cross-section calculation is flatness with respect to \ac{mg} rather than the absolute scale, which can be corrected by a common factor.}
    \label{fig:xsec_comp}
\end{figure}

\section{Technique}
\label{sec:technique}

The physics of \ac{db} has been added to \gf via the creation of an additional physics process class describing the interaction. The major benefit of this technique is compatibility with existing \gf tools. When the \ac{db} process (and in turn the \aprime particle) exist as formal objects in \gf, biasing operators and generic methods to enable and track individual processes can be easily modified to work with the new interaction. The \ac{db} process (``G4DarkBremsstrahlung", documented in Appendix \ref{sec:docs}) is implemented as a discrete electromagnetic process in \gf. 

During each step, \gf uses the total cross section to determine whether a \ac{db} interaction should occur and to calculate the event weight if the process has a bias applied. In this implementation, the cross section is calculated by performing a numerical integration of the \ac{ww} approximation. 
Depending on the lepton and \aprime masses, different variations of the \ac{ww} approximation best follow the \ac{mg} cross section, so three options are implemented in this package. The specifics of the calculation and the different variations are detailed in Appendix \ref{sec:docs:xsec}; the methods best suited to the situations presented are used here.
The result depends on the mass of the \aprime,  the energy of the incident lepton and the material it passes through. Refinement of the total cross section from the \ac{ww} approximation, to ensure that the values match exact tree level calculations, can be done using k-factors during the simulation in \gf \cite{dmg4}.

While the total cross section given by the numerical integral of the \ac{ww} approximation is effective for our purposes, it was found that the differential result it produces has significant deviations from the \ac{mg} results, particularly in the limits of large angular scatters and high energy loss. The kinematics of the interaction are further complicated by the presence of the scattered nucleus. As the nucleus can carry momentum comparable to the lepton, reconstructing the outgoing lepton without including the momentum carried by nucleus produces dramatically different distributions in the recoil momentum and angle. To circumvent these issues the simulation does not construct outgoing leptons by sampling from the differential \ac{ww} cross section, but instead scales  existing \ac{mg} events to the desired incident lepton energy. 

\begin{figure}[tbp]
    \centering
    \includegraphics[width=0.45\textwidth]{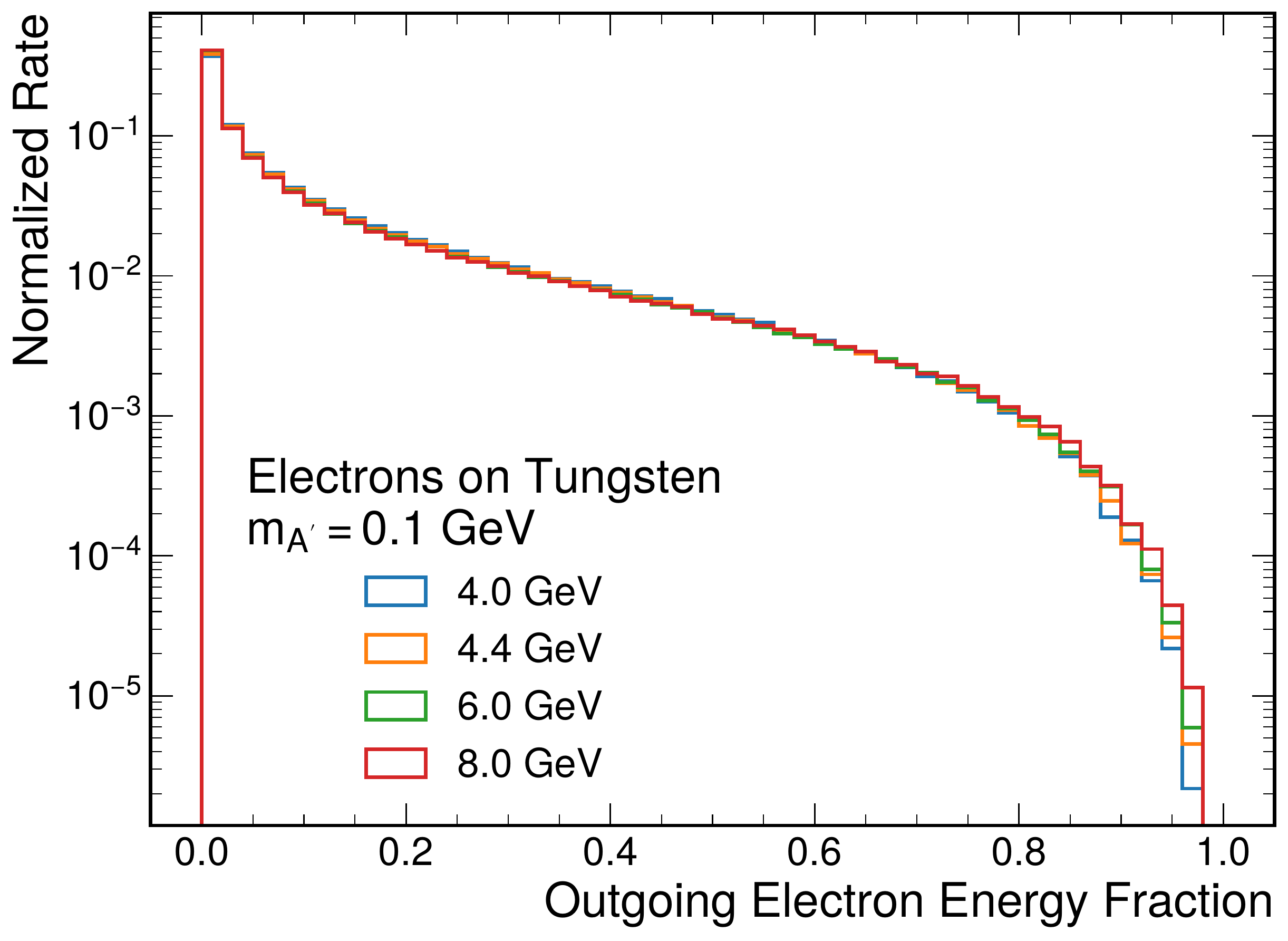}
    \hspace{0.01\textwidth}
    \includegraphics[width=0.45\textwidth]{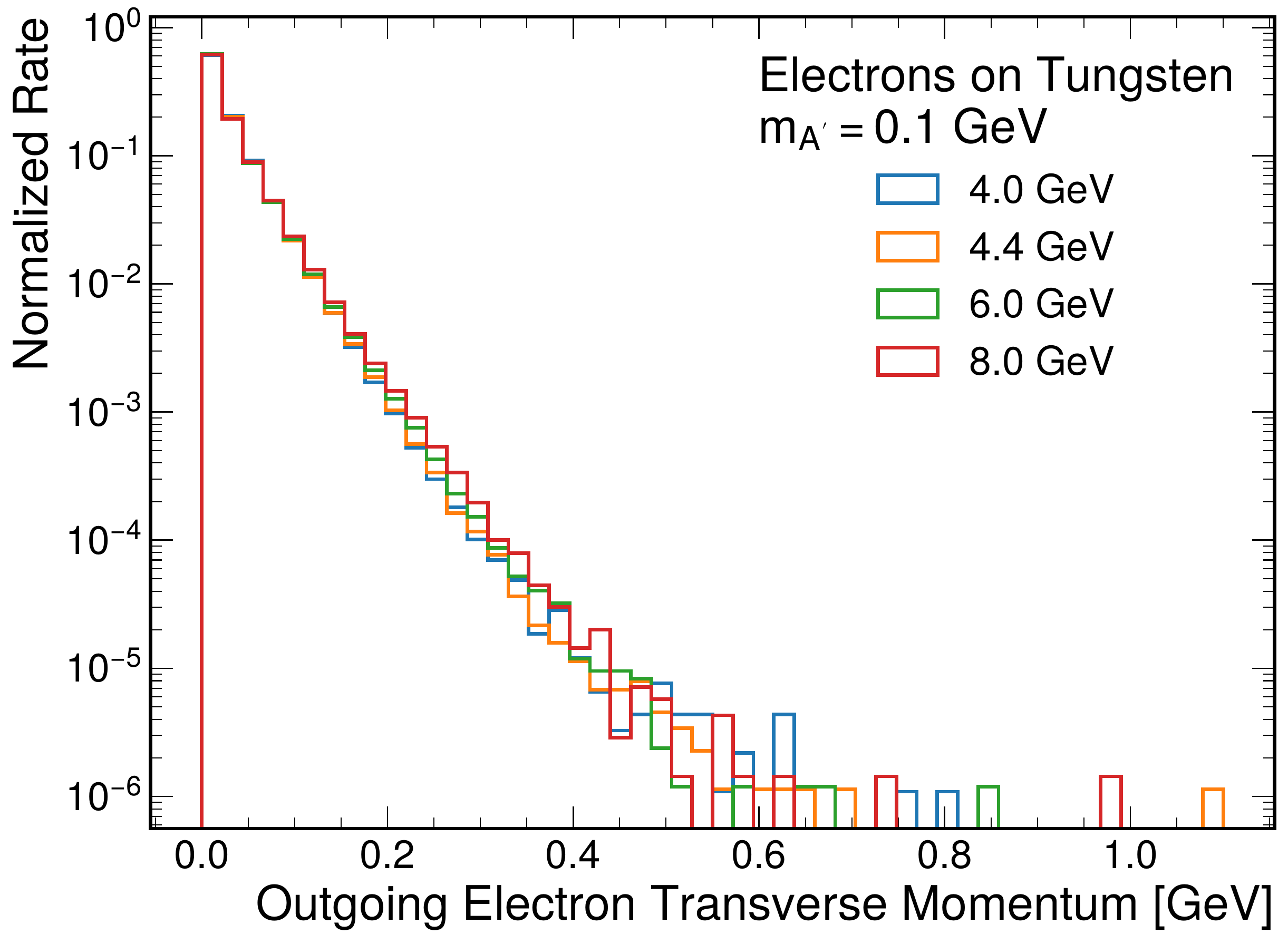}
    \caption{Scattered electron energy fractions and transverse momenta for various energy electrons in Tungsten. The mass of \aprime ($m_{\aprime}$) is taken as 0.1 GeV. The distributions see good agreement for a wide range of incident energies.}
    \label{fig:energies}
\end{figure}

Through study of \ac{mg} events produced over a range of incident lepton energies, two variables with distributions that vary slowly with incident energy were identified: the fraction of the incident kinetic energy imparted to the scattered lepton and the transverse momentum of that lepton with respect to the incident particle (Fig. \ref{fig:energies}). Because their distributions vary slowly with incident energy, we can sample these variables from a set of previously simulated \ac{mg} events at a ``nearby'' energy. The two variables are correlated, so they must be sampled together. The scaling from a sampled \ac{mg} incident energy to the actual \gf incident energy has been found effective over a wide range; nevertheless, the differences become more pronounced as the relative distance between the sampled and true incident energies increases. To reduce inaccuracy introduced during this scaling, a library consisting of pairs of these variables at a variety of incident energies and target materials can be loaded. The impact of these variations can be limited by using finer-binned energies in the sampled \ac{mg} libraries (e.g.~at energies spaced by 10\% over the range where greatest precision is required). 

When \gf determines that a \ac{db} event should occur, the process selects a random event from the nearest available lepton energy and target material in the loaded \ac{mg} library. The simulation then creates a new lepton with the same transverse momentum as the sampled event and scales the outgoing kinetic energy to match the fraction of incident kinetic energy of the sampled lepton, then randomly selects the azimuthal angle with respect to the initial lepton momentum. In the rare case where the total energy of the new lepton is smaller than the selected transverse momentum, alternate events will be sampled until one is found with a scaled final energy larger than its transverse momentum. 

The simulation assumes that the scattered nucleus does not receive sufficient momentum to produce an observable signal in the detector sensitive volumes; therefore, no \gf particle for the scattered nucleus is created and, as a result, energy and momentum cannot be precisely conserved using only the \aprime and scattered lepton. The simulation produces an \aprime using conservation of momentum only so that the incident lepton can be reconstructed from simulation-level information about the \aprime and the scattered lepton. No physics processes have been implemented for the emitted \aprime. A study which depends on the kinematic distributions of the \aprime may require further extension of this technique.

\begin{figure}[tbp]
    \centering
    \includegraphics[width=0.45\textwidth]{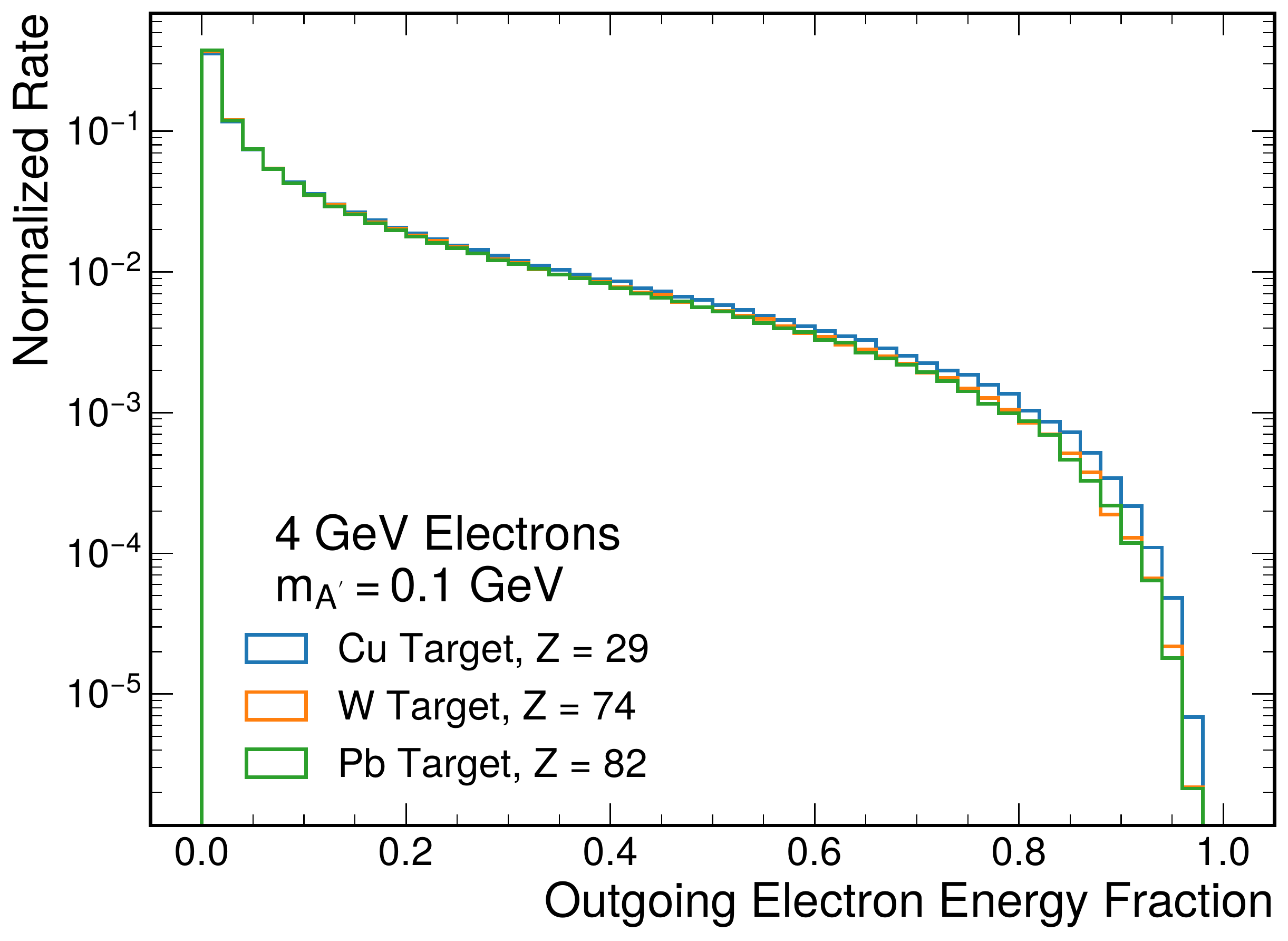}
    \hspace{0.01\textwidth}
    \includegraphics[width=0.45\textwidth]{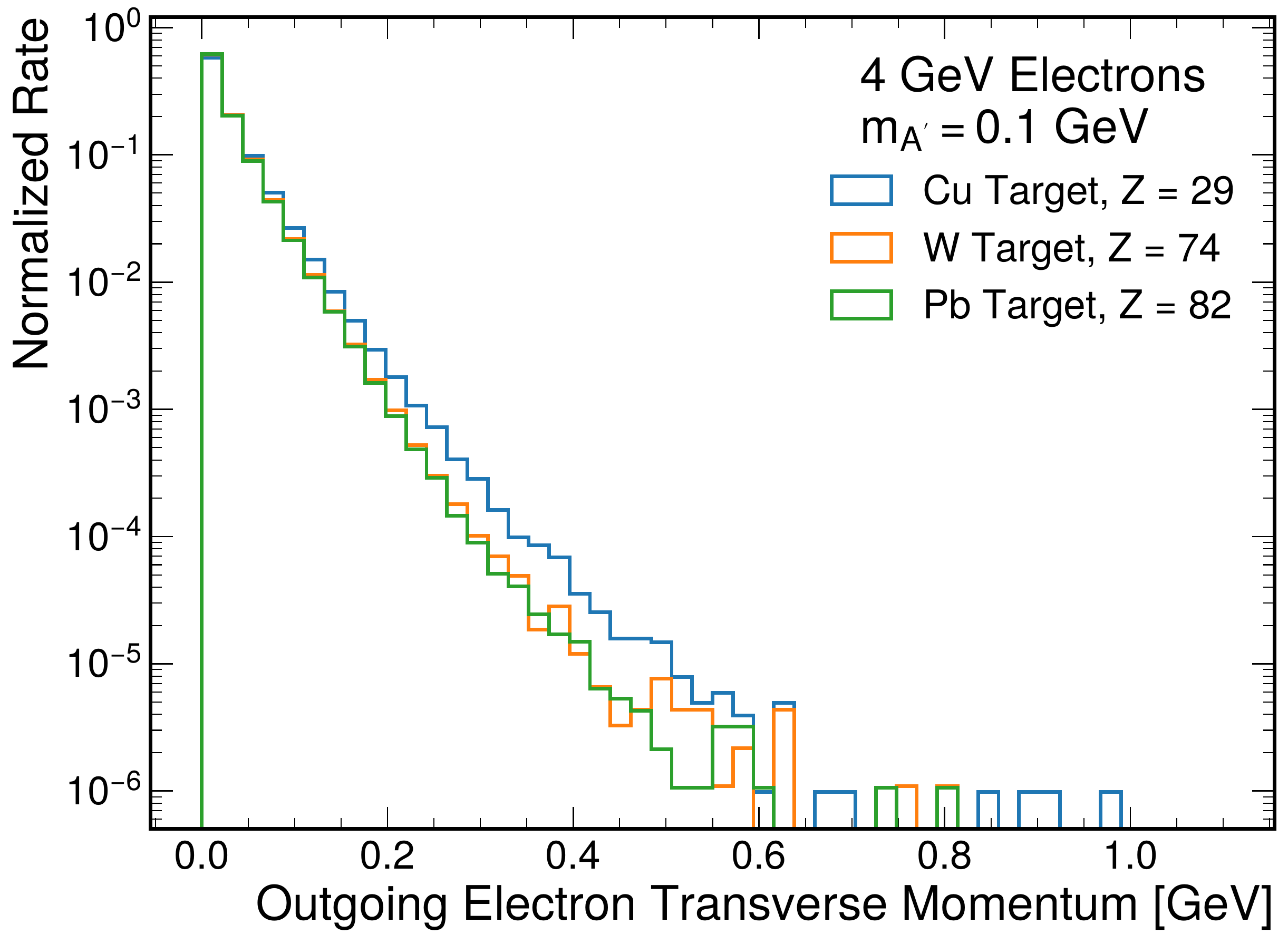}
    \par
    \includegraphics[width=0.45\textwidth]{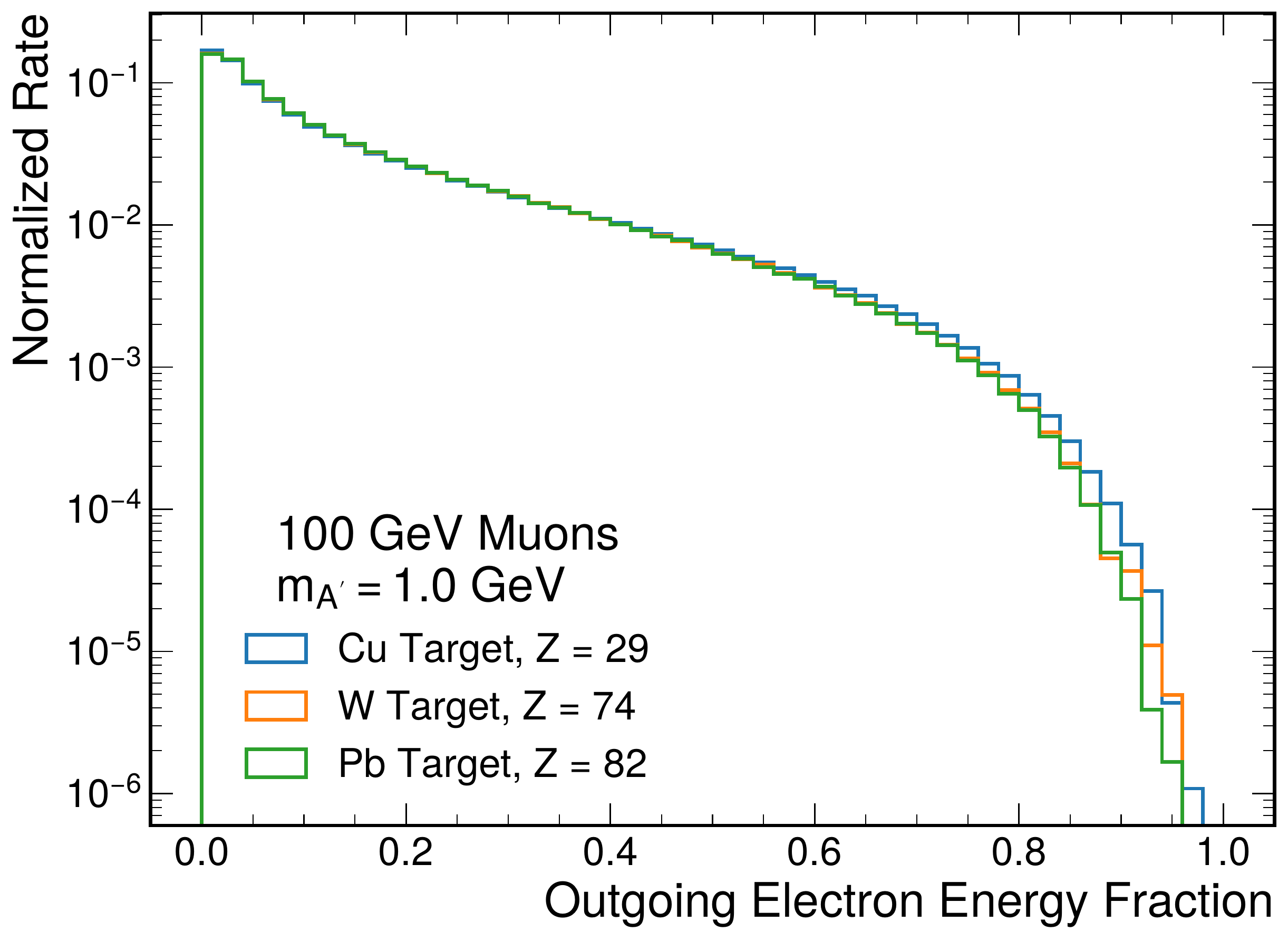}
    \hspace{0.01\textwidth}
    \includegraphics[width=0.45\textwidth]{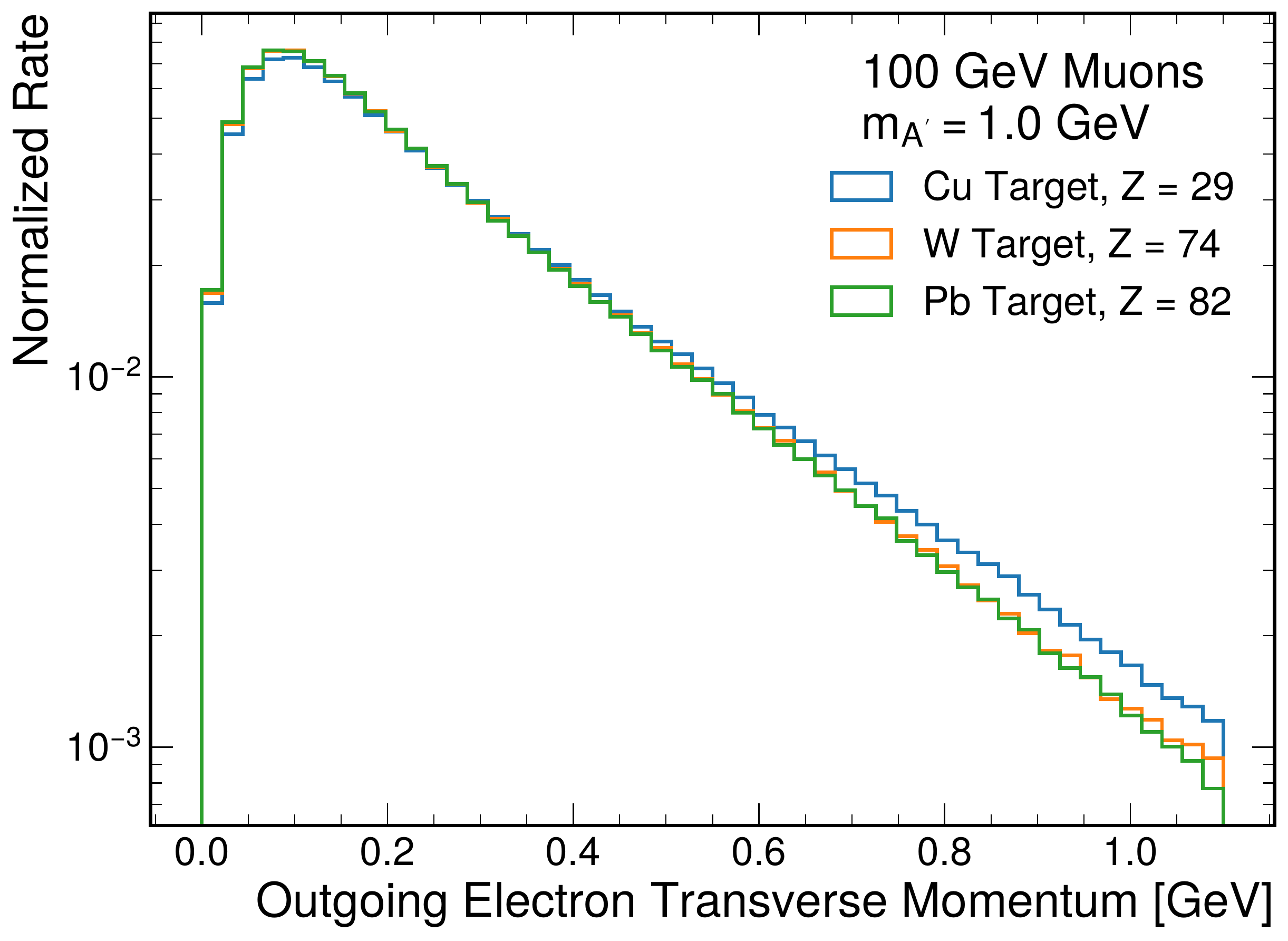}
    \caption{Scattered lepton energy fractions and transverse momenta for 4.0 GeV electrons and 100.0 GeV muons in Copper, Tungsten, and Lead. The mass of \aprime ($m_{\aprime}$) is taken as 0.1 GeV for electrons and 1.0 GeV for muons. The distributions see good agreement for the small change in Z from Lead to Tungsten, and begin to differ with the large change to Copper.}
    \label{fig:mats}
\end{figure}

The kinematics of the outgoing lepton may depend weakly on the composition of the target material. The core assumptions of the scaling technique, slow variation of transverse momentum and outgoing energy fraction spectra with respect to incident energy, are found to remain valid across a wide range of target materials while the underlying distributions of those variables change weakly (Fig.~\ref{fig:mats}). In composite materials, the simulation of the outgoing kinematics can randomly select a initiating atom from the target, weighted by the cross section and material abundance, in order to sample from the correct kinematics. Additionally, the outgoing energy and angle distributions are very similar for materials with small changes in Z, and in materials with many constituents the process can sample from similar nuclei without needing to generate entries in the \ac{mg} library for every potential initiating nucleus. 

\begin{figure}[!htb]
    \centering
    \includegraphics[width=0.45\textwidth]{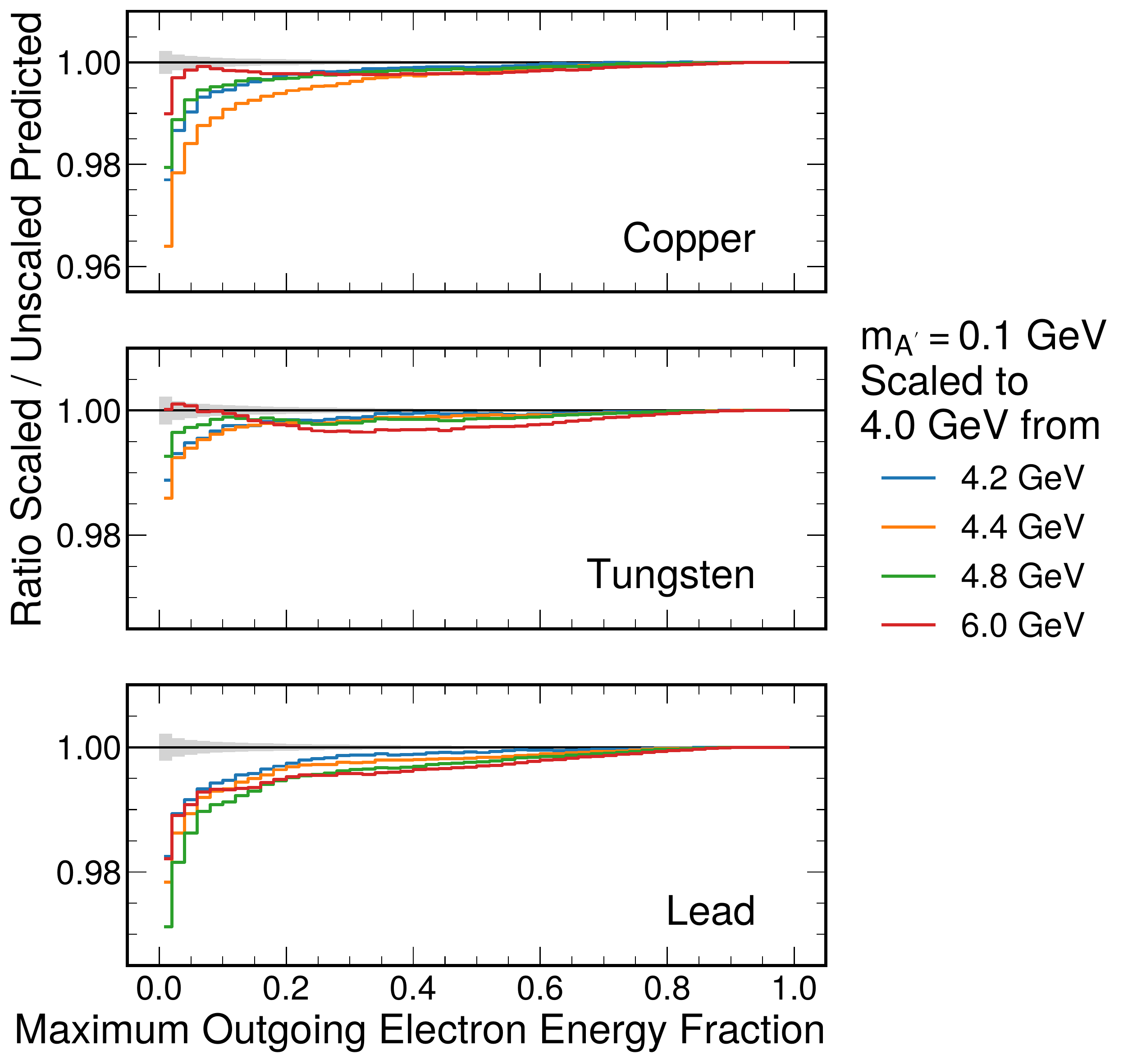}
    \hspace{0.01\textwidth}
    \includegraphics[width=0.45\textwidth]{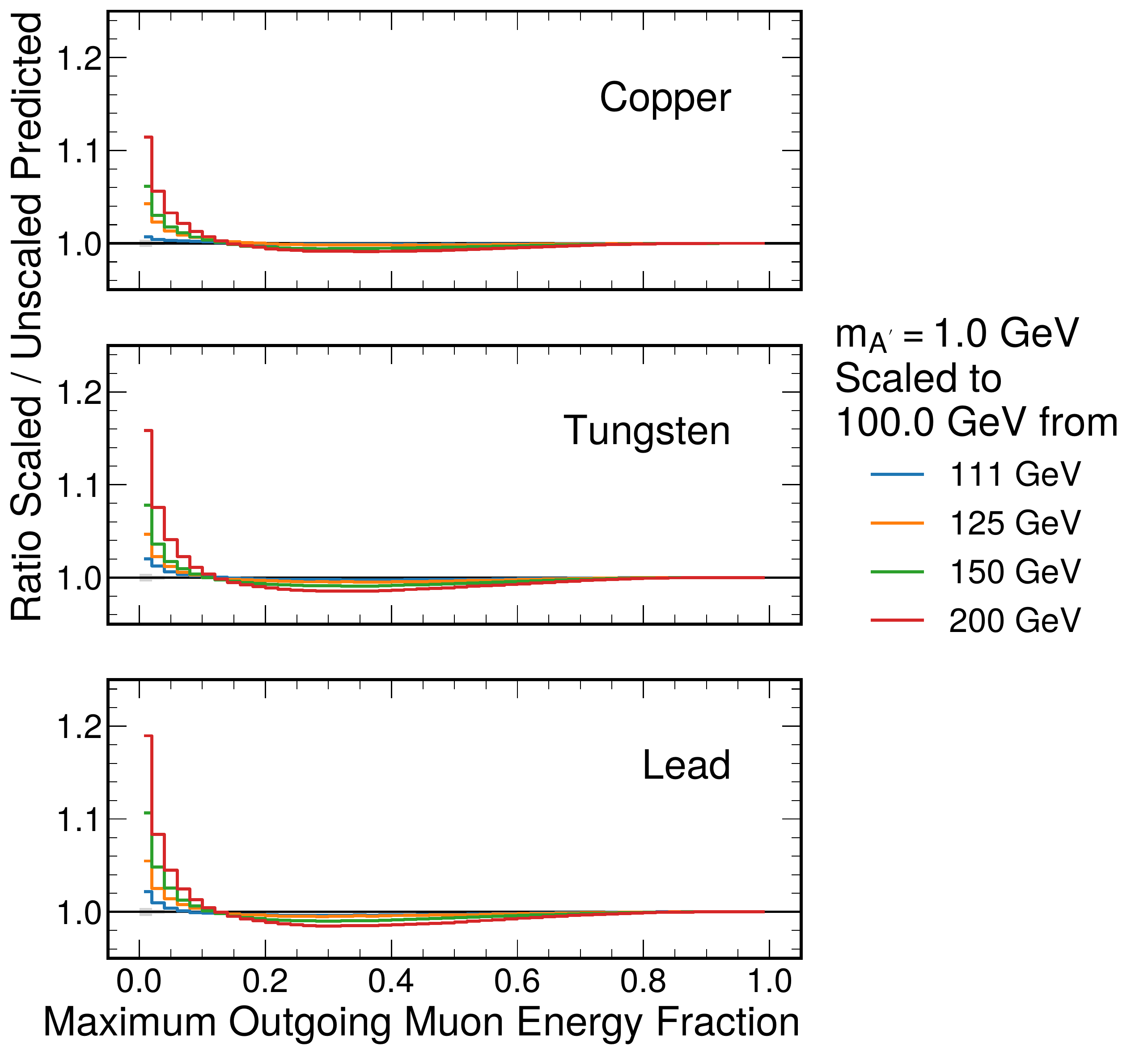}
    \caption{Comparison of the predicted events to un-scaled \ac{mg} with a certain maximum outgoing lepton energy for electrons scaled to 4.0 GeV and muons scaled to 100 GeV in Copper, Tungsten, and Lead. The mass of \aprime ($m_{\aprime}$) is taken as 0.1 GeV for the electron examples and 1 GeV for the muon examples.}
    \label{fig:scalingE}
\end{figure}
\begin{figure}[!htb]
    \centering
    \includegraphics[width=0.45\textwidth]{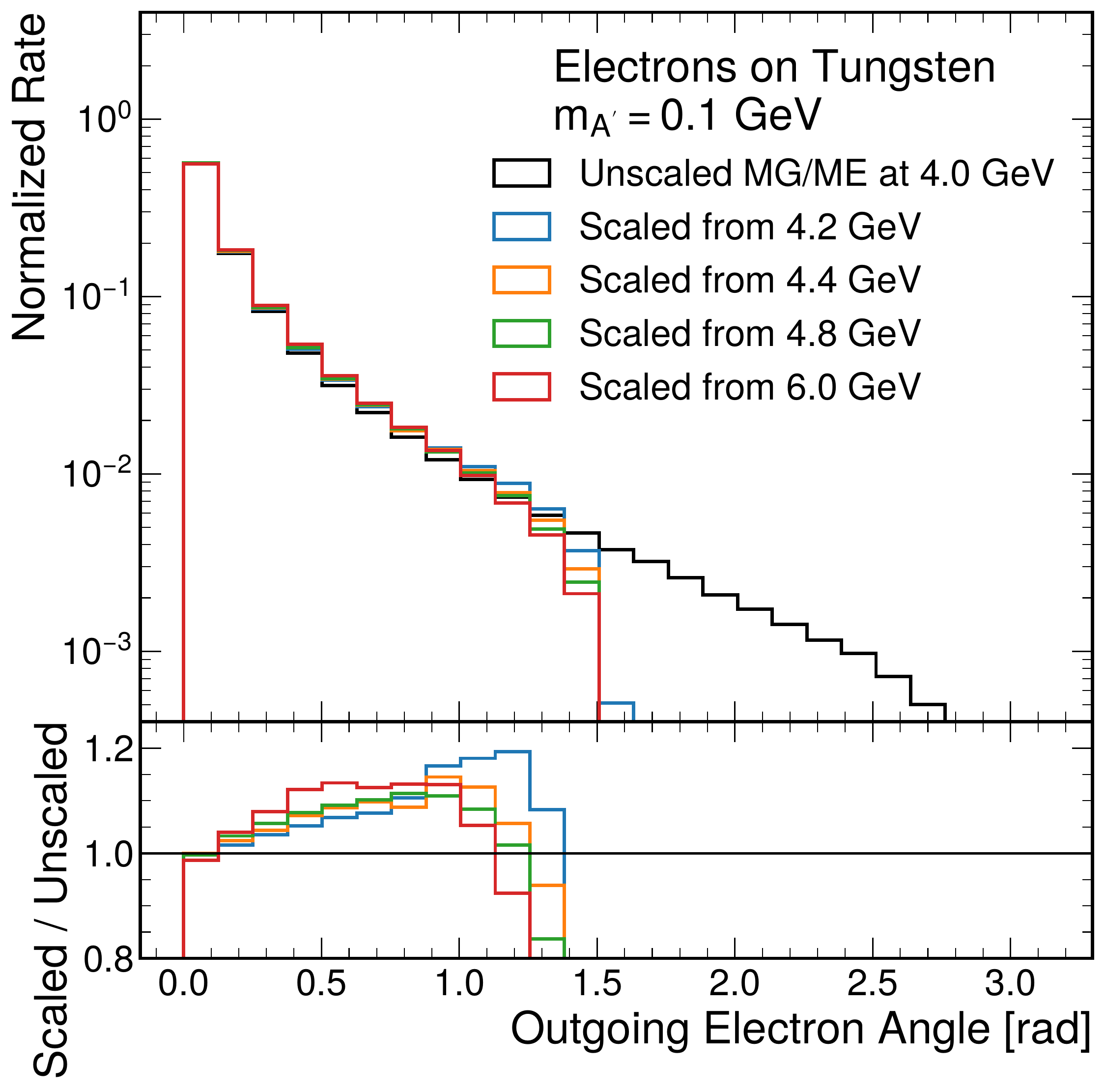}
    \hspace{0.01\textwidth}
    \includegraphics[width=0.45\textwidth]{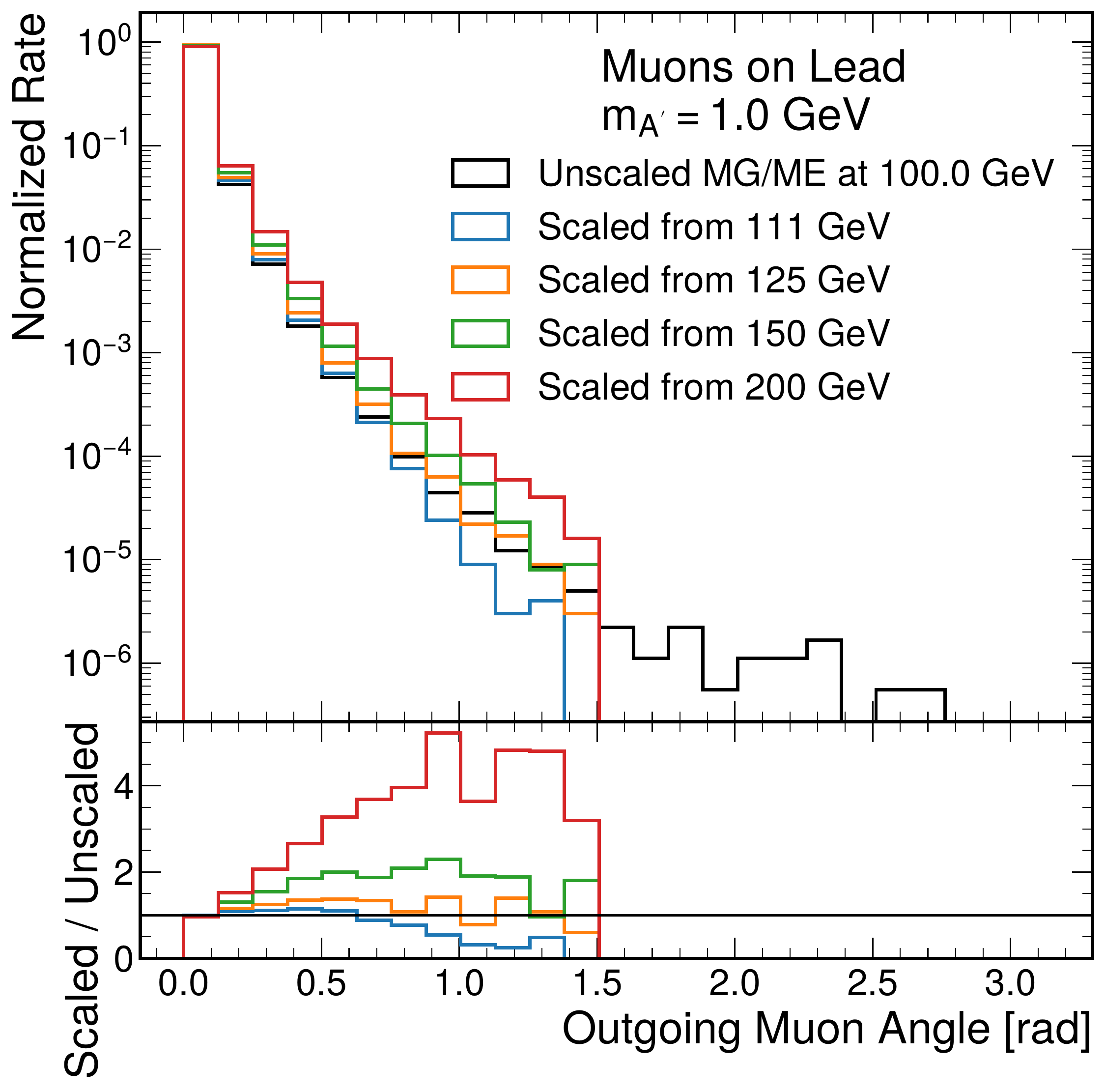}
    \caption{Lepton deflection angle for electrons scaled to 4.0 GeV in Tungsten and muons scaled to 100 GeV in Lead compared to un-scaled \ac{mg}. The mass of \aprime ($m_{\aprime}$) is taken as 0.1 GeV for the electron example and 1 GeV for the muon example.}
    \label{fig:scalingAngle}
\end{figure}

\section{Validation}
\label{sec:validation}

Comparing the recoil lepton kinematics produced using this scaling procedure to a sample of \ac{mg} events at the target incident energy shows that this procedure does not significantly distort the distributions (Figures \ref{fig:scalingE} and \ref{fig:scalingAngle}). Moreover, these figures show that the scaling procedure more-closely resembles the ``true" \ac{mg} distribution as the incident energy the procedure scales from becomes closer to the target incident energy.

\newpage

The outgoing lepton angle (Fig. \ref{fig:scalingAngle}) is less well reconstructed using this procedure; however, the distributions stay within roughly 10\% of the \ac{mg} distributions when the difference between sampled incident energy and target incident energy is kept below 10\%. This procedure \emph{does not} reconstruct the sign of the outgoing lepton's z-momentum, so it is always chosen to be positive. This choice produces the only large deviation between the scaling procedure and \ac{mg} that is at a low ($< 1$\%) relative rate and consists of backwards-scattered leptons which could be expected to be removed by downstream analysis cuts.

\section{Reach}
\label{sec:reach}

This \ac{db} simulation technique can be used to allow the \ac{db} process to occur throughout material volumes within a \gf simulation. Another software package that integrates this process into \gf is \ac{dmg4} which offers a wide range of dark sector processes in addition to the \ac{db} process. Both \ac{dmg4} and the technique described here (\ac{g4db}) offer an extension of \gf for studying the behavior of new physics within experiments' detector volumes. To demonstrate what can be learned from this technique, the behavior of both \ac{dmg4} and \ac{g4db} are studied within four situations: $4$~GeV electron and $100$~GeV muon beams are fired on ``thin" and ``thick" targets of two different materials using scalings of 10\%.

\begin{figure}[!htb]
    \centering
    \begin{subfigure}[b]{0.45\textwidth}
        \centering
        \includegraphics[width=\textwidth]{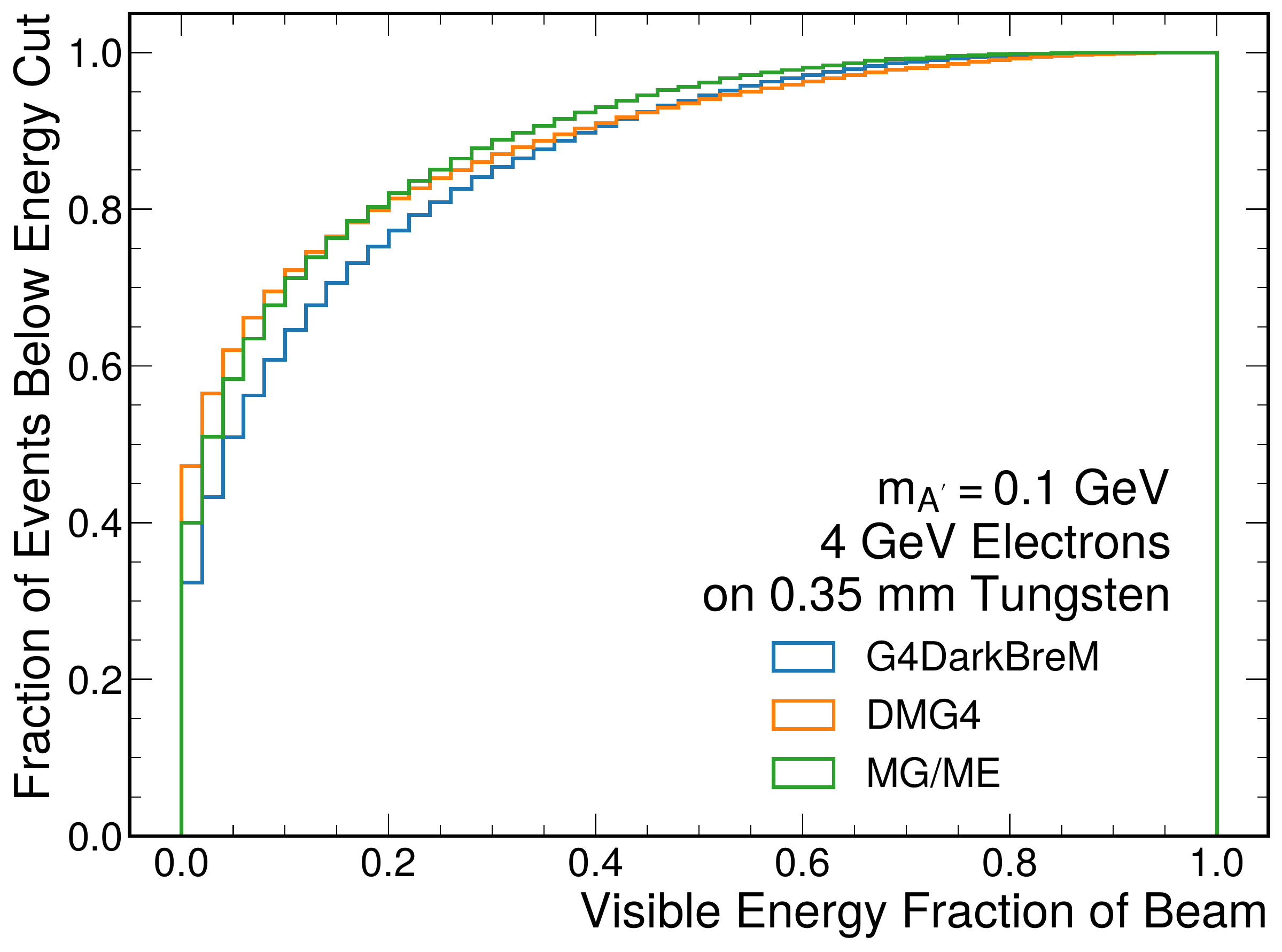}
    \end{subfigure}
    \begin{subfigure}[b]{0.45\textwidth}
        \centering
        \includegraphics[width=\textwidth]{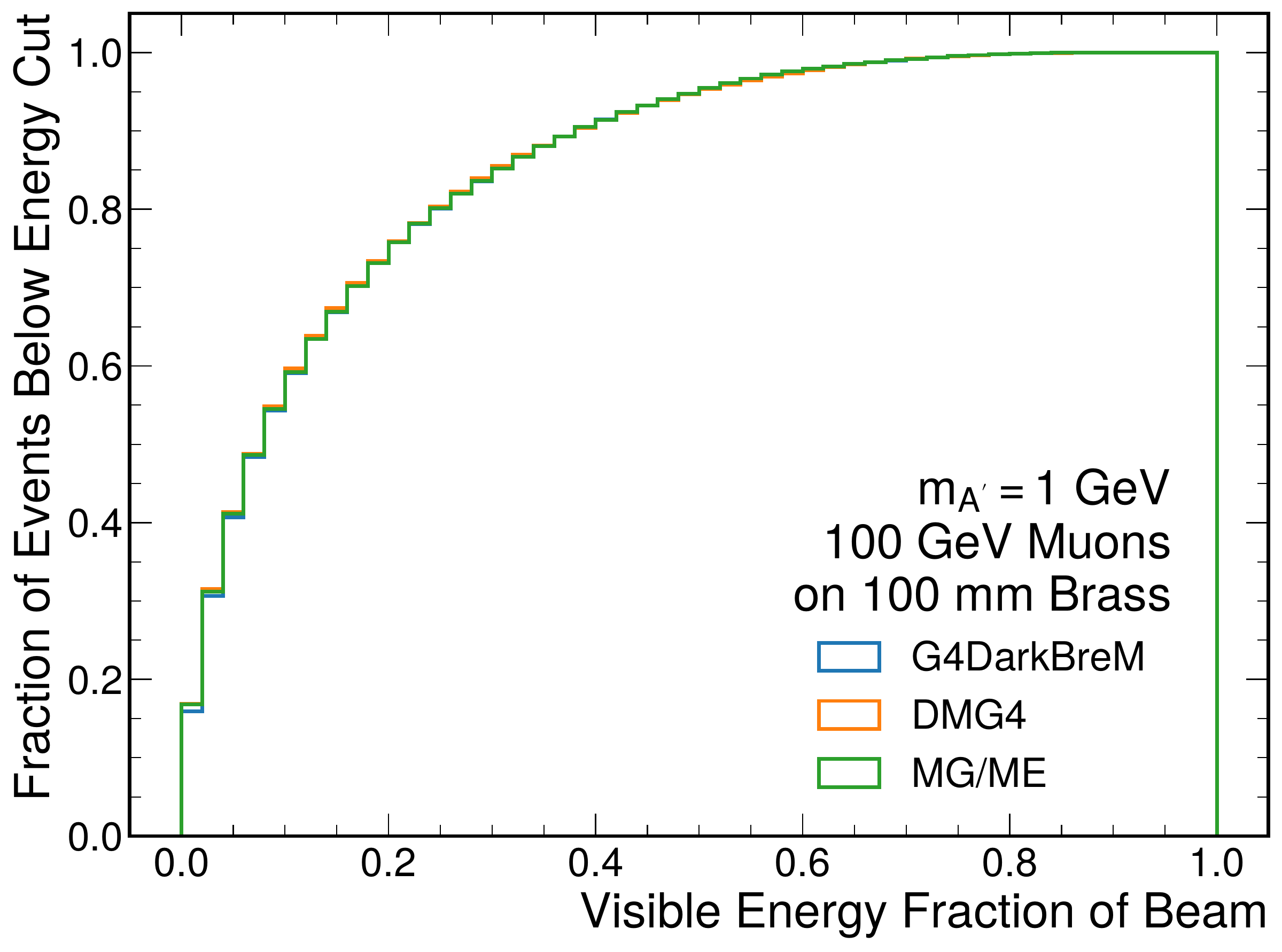}
    \end{subfigure}
    \caption{Comparison of \ac{dmg4} and \ac{g4db} visible energy distributions in thin targets. In the thin-target limit, using a fixed incident energy approximation is valid, so the un-scaled \ac{mg} distributions are expected to be close. Shown are $4~$GeV electrons incident on $0.35$~mm of tungsten with the mass of \aprime ($m_{A'}$) taken as $0.1$~GeV and $100~$GeV muons incident on $100$~mm of brass with the mass of \aprime ($m_{\aprime}$) taken as $1.0$~GeV.}
    \label{fig:thin_target_visible_energy}
\end{figure}
\begin{figure}[!htb]
    \centering
    \begin{subfigure}[b]{0.45\textwidth}
        \centering
        \includegraphics[width=\textwidth]{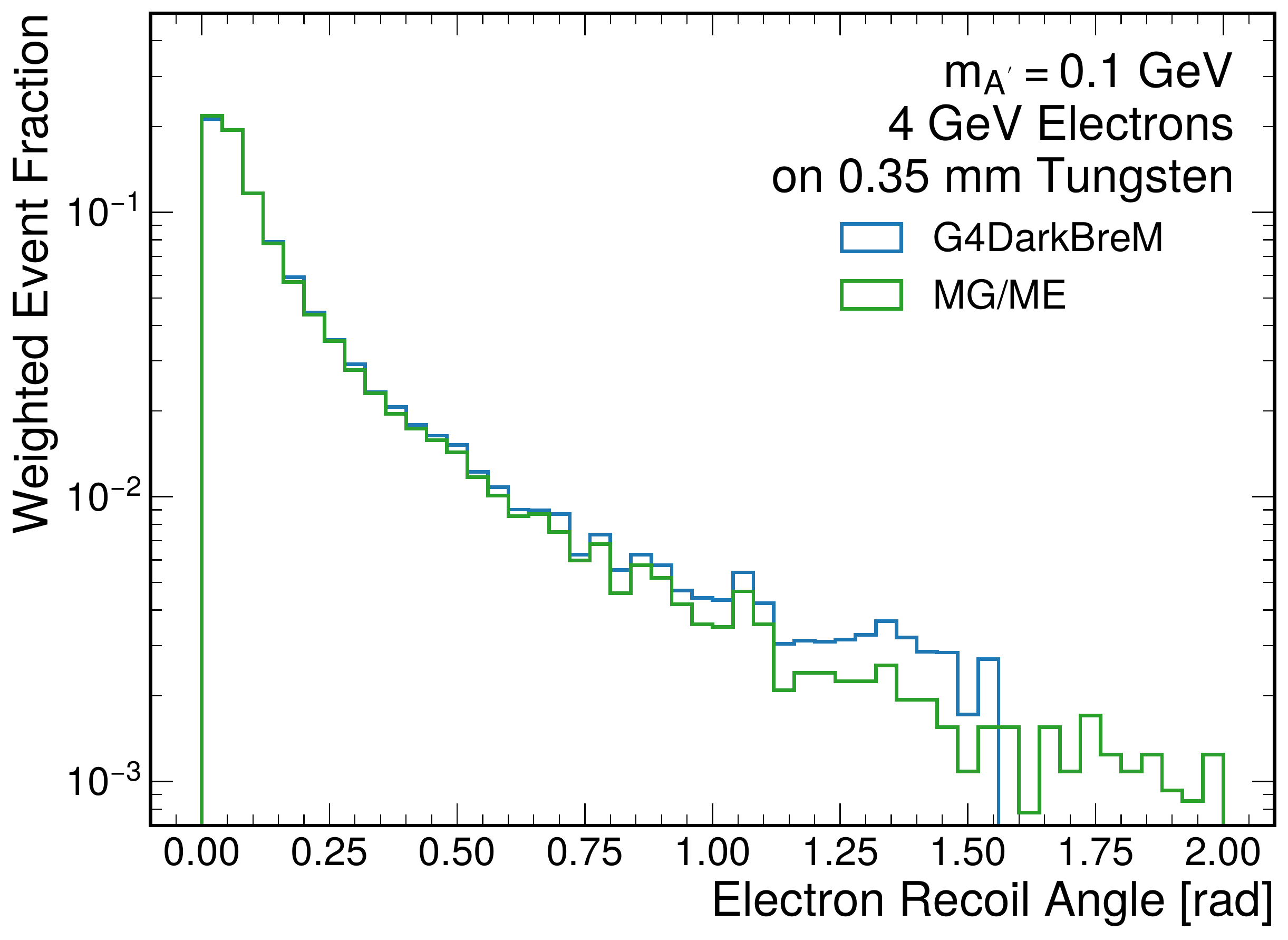}
    \end{subfigure}
    \begin{subfigure}[b]{0.45\textwidth}
        \centering
        \includegraphics[width=\textwidth]{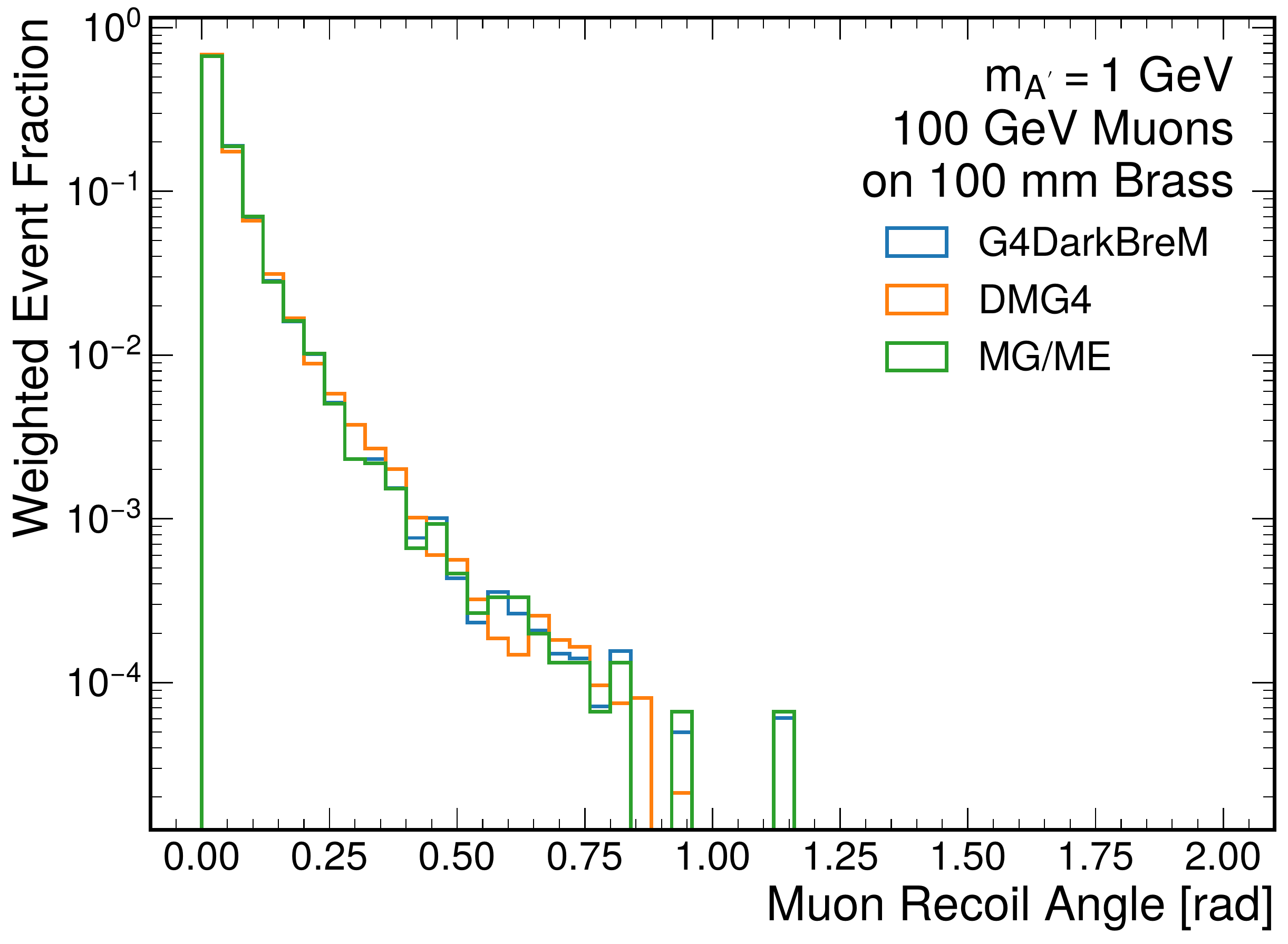}
    \end{subfigure}
    \caption{Comparison of \ac{dmg4} and \ac{g4db} recoil angle distributions in thin targets. In the thin-target limit, using a fixed incident energy approximation is valid, so the un-scaled \ac{mg} distributions are expected to be close. Shown are $4~$GeV electrons incident on $0.35$~mm of tungsten with the mass of \aprime ($m_{A'}$) taken as $0.1$~GeV and $100~$GeV muons incident on $100$~mm of brass with the mass of \aprime ($m_{\aprime}$) taken as $1.0$~GeV. \ac{dmg4} is omitted in the electron case since it does not implement a recoil angle model for electrons.}
    \label{fig:thin_target_recoil_angle}
\end{figure}
\begin{figure}[!htb]
    \centering
    \begin{subfigure}[b]{0.45\textwidth}
        \centering
        \includegraphics[width=\textwidth]{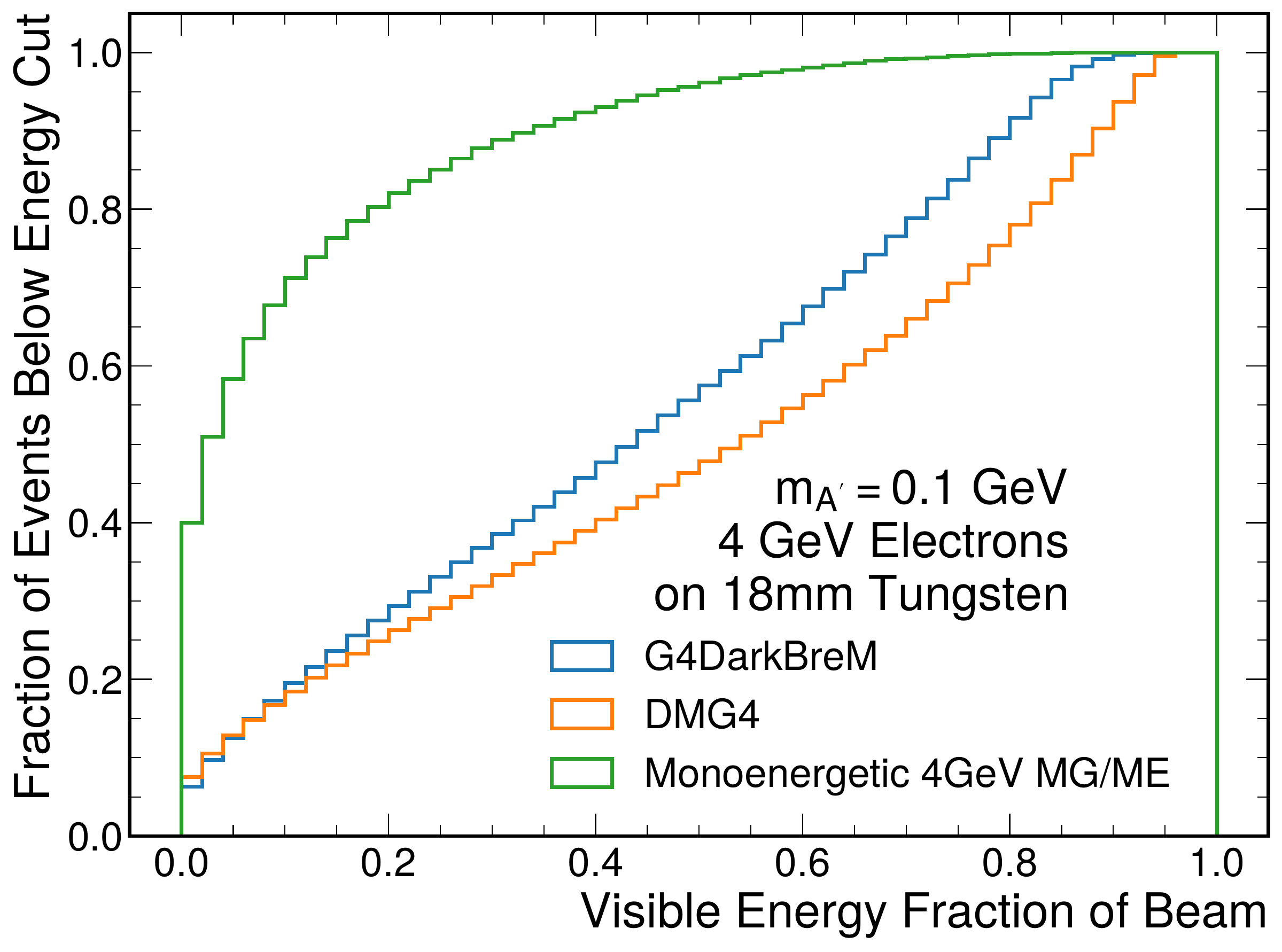}
    \end{subfigure}
    \begin{subfigure}[b]{0.45\textwidth}
        \centering
        \includegraphics[width=\textwidth]{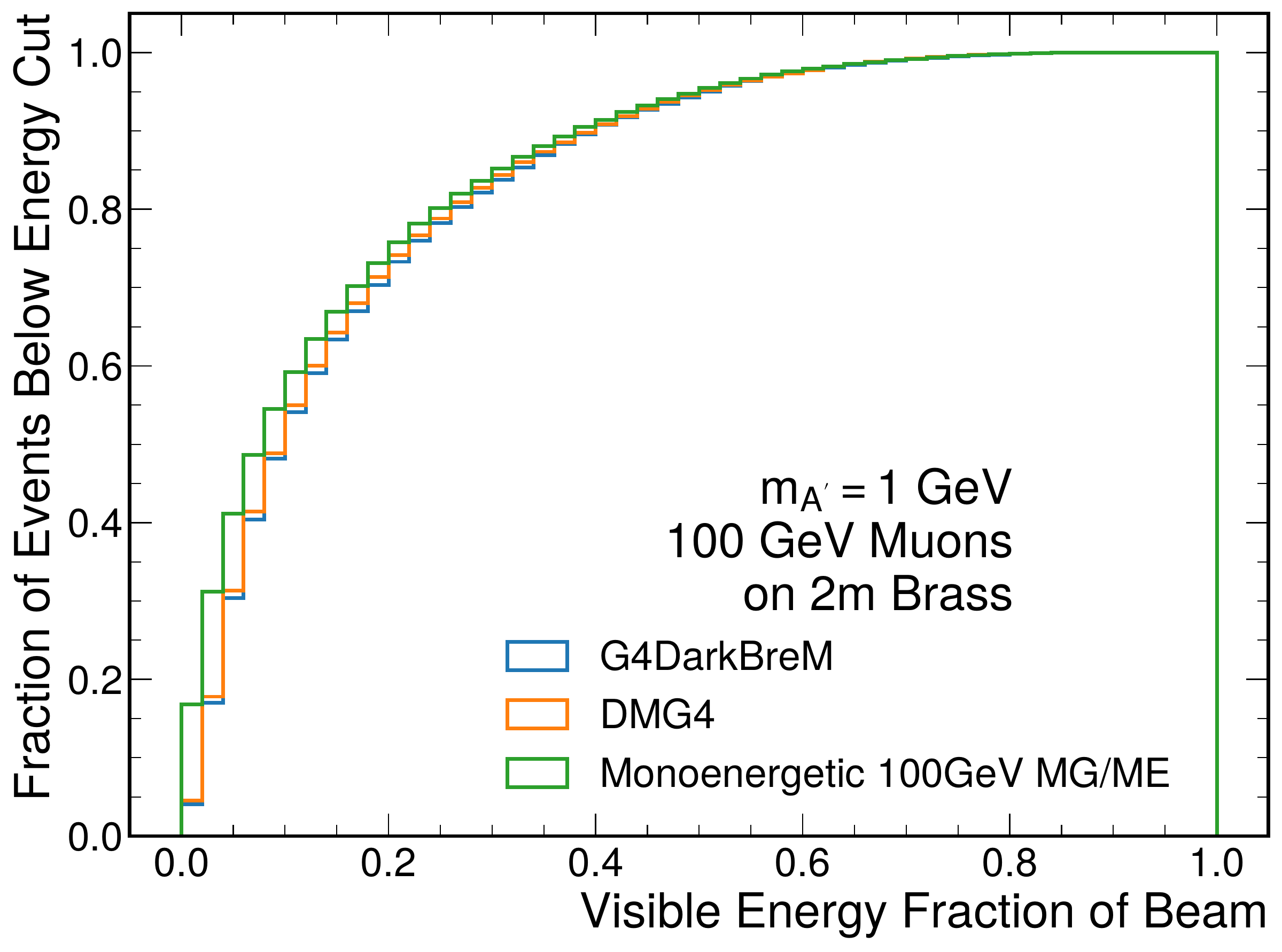}
    \end{subfigure}
    \caption{Comparison of \ac{dmg4} and \ac{g4db} visible energy distributions in thick targets for $4~$GeV electrons incident on $18$~mm of tungsten with the mass of \aprime ($m_{A'}$) taken as $0.1$~GeV and $100~$GeV muons incident on $2$~m of brass with the mass of \aprime ($m_{\aprime}$) taken as $1.0$~GeV.}
    \label{fig:thick_target_visible_energy}
\end{figure}
\begin{figure}[!htb]
    \centering
    \begin{subfigure}[b]{0.45\textwidth}
        \centering
        \includegraphics[width=\textwidth]{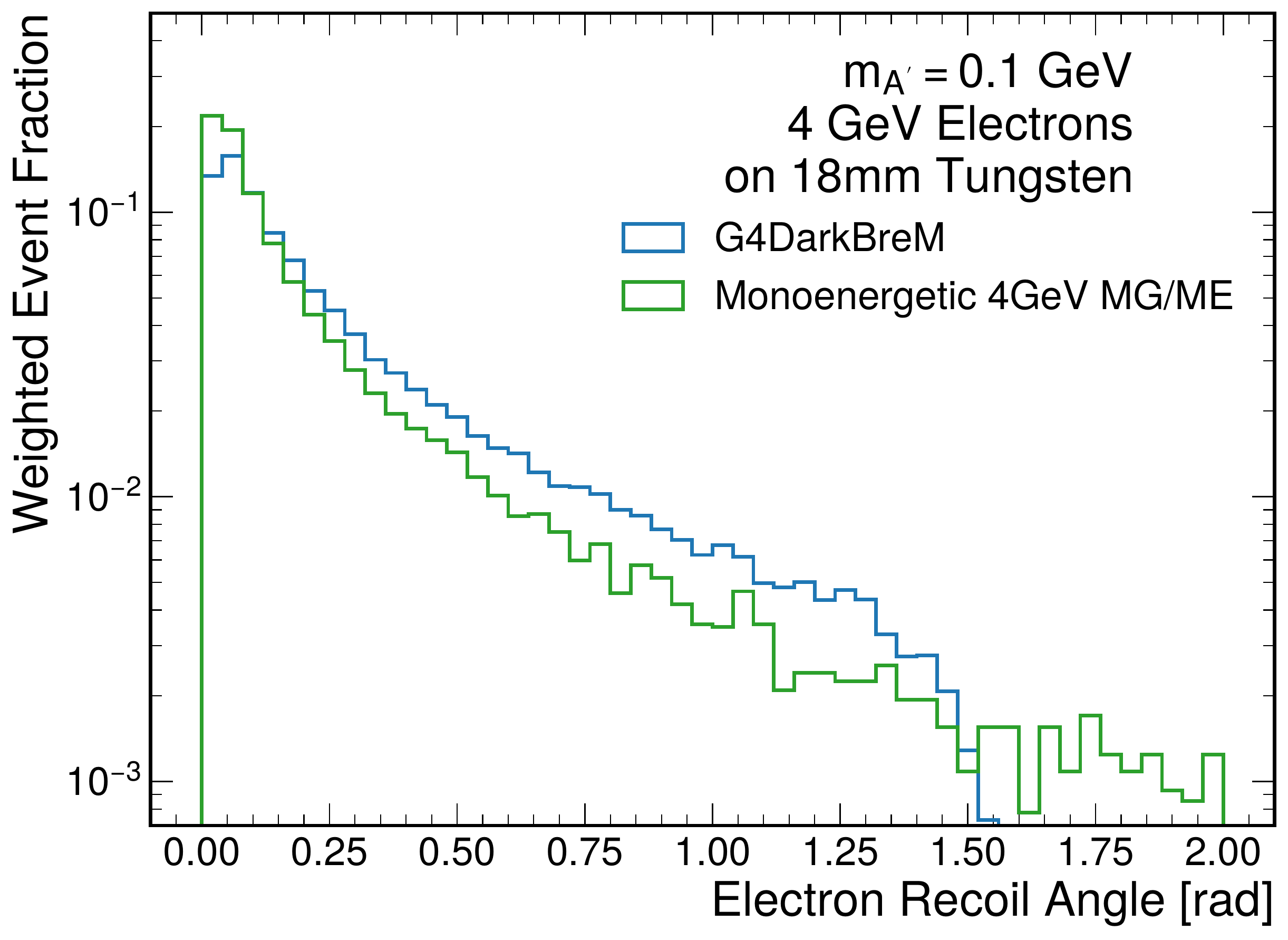}
    \end{subfigure}
    \begin{subfigure}[b]{0.45\textwidth}
        \centering
        \includegraphics[width=\textwidth]{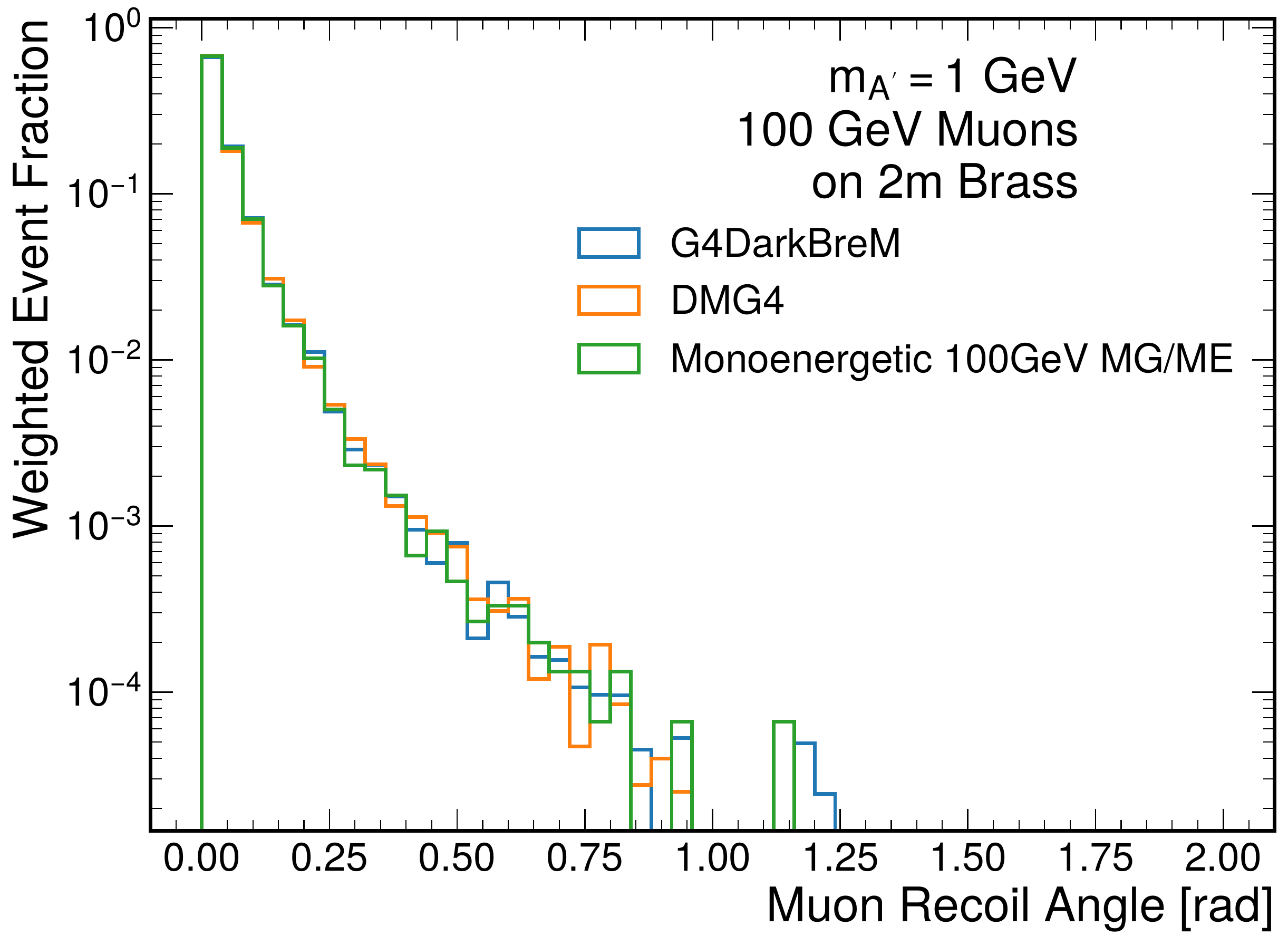}
    \end{subfigure}
    \caption{Comparison of \ac{dmg4} and \ac{g4db} recoil angle distributions in thick targets for $4~$GeV electrons incident on $18$~mm of tungsten with the mass of \aprime ($m_{A'}$) taken as $0.1$~GeV and $100~$GeV muons incident on $2$~m of brass with the mass of \aprime ($m_{\aprime}$) taken as $1.0$~GeV. \ac{dmg4} is omitted in the electron case since it does not implement a recoil angle model for electrons.}
    \label{fig:thick_target_recoil_angle}
\end{figure}

The largest benefit of embedding the \ac{db} process into \gf is that the potential observables that experiments can measure are more readily calculated. The angle of the recoiling lepton (here defined relative to the beam direction) and the total visible energy (i.e. from \ac{sm} particles) can be extracted from the \gf simulation in a similar way as standard processes already within \gf.

In the thin target, a vast majority of beam leptons approximately maintain the beam energy up until the \ac{db} interaction; therefore, observables are expected to closely follow the \ac{mg} distributions which maintain the assumption that the lepton undergoing the \ac{db} interaction has exactly the beam energy. Both \ac{dmg4} and \ac{g4db} match this expectation fairly well in the fraction of visible energy (Fig. \ref{fig:thin_target_visible_energy}). \ac{g4db} is able to follow this assumption for the recoiling angle of both electrons and muons (Fig. \ref{fig:thin_target_recoil_angle}) effectively replicating the validation plots already presented. The small deviations between \ac{mg} and \ac{g4db} in these samples can be attributed to standard processes occurring \emph{before} the \ac{db} interaction within the \gf simulation.

The thick target examples offer a glimpse at the power of the \ac{g4db} technique. Here, we can maintain connection with \ac{mg} distributions at a variety of incident energies while allowing the standard \gf interactions to take place both before and after the \ac{db} process of interest. In this case, both visible energy distributions (Fig. \ref{fig:thick_target_visible_energy}) flatten due to energy losses occurring before the lepton undergoes the \ac{db}. This effect is particularly strong in the electron case where it is common for the beam electron to begin a standard electromagnetic shower and the \ac{db} may occur with a secondary electron at a much lower energy (the thick electron target is $\sim 5 X_0$). Muons only lose an average of $\sim 0.1\%$ of their kinetic energy to standard electromagnetic processes in $2$m of brass so this distribution is closer to the monoenergetic \ac{mg} distribution.
Likewise, the recoil-lepton angle distributions (Figure \ref{fig:thick_target_recoil_angle}) show a broadening mostly caused by upstream interactions perturbing the incident lepton off the beam axis. Again, both \ac{dmg4} and \ac{g4db} have similar distributions for the visible energy observable; however, only \ac{g4db} provides physical recoil angle distributions for electrons.

\section{Conclusion}
A novel technique for simulating \ac{db} has been implemented and shown to not only faithfully replicate the distributions generated by \ac{mg} but also to extend this simulation to use cases where the incident lepton energy can vary widely often caused by other interactions before the \ac{db} of interest. This implementation into \gf gives users the ability to treat the \ac{db} process like any other standard process in the simulation of their experiment, making it easier to study the intricate differences between signal and backgrounds.

\newpage

\bibliographystyle{elsarticle-num}
\bibliography{references}

\newpage

\begin{appendices}

\section{Implementation of Dark Brem Process}
\label{sec:docs}
The \ac{g4db} method functions by writing a \texttt{G4VDiscreteProcess}. The \ac{db} process \texttt{G4DarkBremsstrahlung} has two important handles into the simulation that it inherits from \texttt{G4VDiscreteProcess}.
\begin{enumerate}
    \item \texttt{GetMeanFreePath} --- calculate an effective mean free path given the current state of the particle
    \item \texttt{PostStepDoIt} --- actually perform the \ac{db} interaction
\end{enumerate}
These are the two core additional methods; and the ones that require the most complexity. In order to separate out these core, complex methods from the other process methods that are required to interface with \gf, the cross section calculations and \ac{db} simulation is done in a ``Dark Brem Model" which is owned by the process. The class \texttt{G4DarkBremsstrahlungModel} is an abstract class detailing the necessary methods for defining a specific \ac{db} model. A specific model \texttt{G4DarkBreMModel} is defined and uses an imported library of \ac{db} events to simulate the outgoing kinematics of a \ac{db} interaction, scaling them as described in this note. This design choice was implemented with the assumption that future, alternative \ac{db} models can be implemented (e.g. integrating an event generator into \gf without using LHE files to transfer the information). In Table \ref{tab:implementation_classes} is a list of classes associated with the \ac{db} process and their intended use.
    
As mentioned before, the main developments are in \texttt{G4DarkBreMModel}. This class takes an \aprime mass and a directory of LHE files of \ac{db} events as well as some optional parameters that allow the user to tune how it operates. When \gf decides for an lepton to undergo a \ac{db}, the method \texttt{G4DarkBreM::GenerateChange} is called. In this method, the model follows the scaling procedure defined in Section \ref{sec:technique}. Then the outgoing particles (an \aprime and a significantly reduced lepton) are added as \emph{new} particles to \gf, while the primary lepton is stopped.

The ``Scaling Method" parameter is more complicated and is given more detail in Appendix \ref{sec:docs:method}. There are several other small changes and additions that are required for the integration of a new user-defined process into \gf, but none of them are physics related. Many are mainly there to allow the user to pass the parameters to the model through the \texttt{RunManager}.

\begin{table}[!htb]
    \centering
    \begin{tabular}{|c|p{0.5\textwidth}|}
        \hline
        Class & Purpose \\
        \hline
        \texttt{G4DarkBremsstrahlung}      &
        A process registering and configuring the dark brem model
        \\
        \texttt{ElementXsecCache} &
        A class that caches cross sections calculated by a model
        \\
        \texttt{G4DarkBremsstrahlungModel} &
        A model that performs the dark brem
        \\
        \texttt{G4DarkBreMModel} &
        A specific model that uses an imported dark brem vertex library
        \\
        \texttt{G4APrime}                   &
        The particle definition for the dark photon A'
        \\
        \hline
    \end{tabular}
    \caption{A list of the class additions that allow for \gf{} to handle dark brem in a variety of ways.}
    \label{tab:implementation_classes}
\end{table}

\subsection{Total Cross Section Calculation}
\label{sec:docs:xsec}

The estimate for the total cross section given the material and the lepton's energy is done using a \ac{ww} approximation. The \ac{ww} approximation is done using Boost's Math Quadrature library to numerically calculate the integrals. The formulas are listed here for reference.

\begin{figure}[!htb]
    \centering
    \begin{subfigure}[b]{0.45\textwidth}
        \centering
        \includegraphics[width=\textwidth]{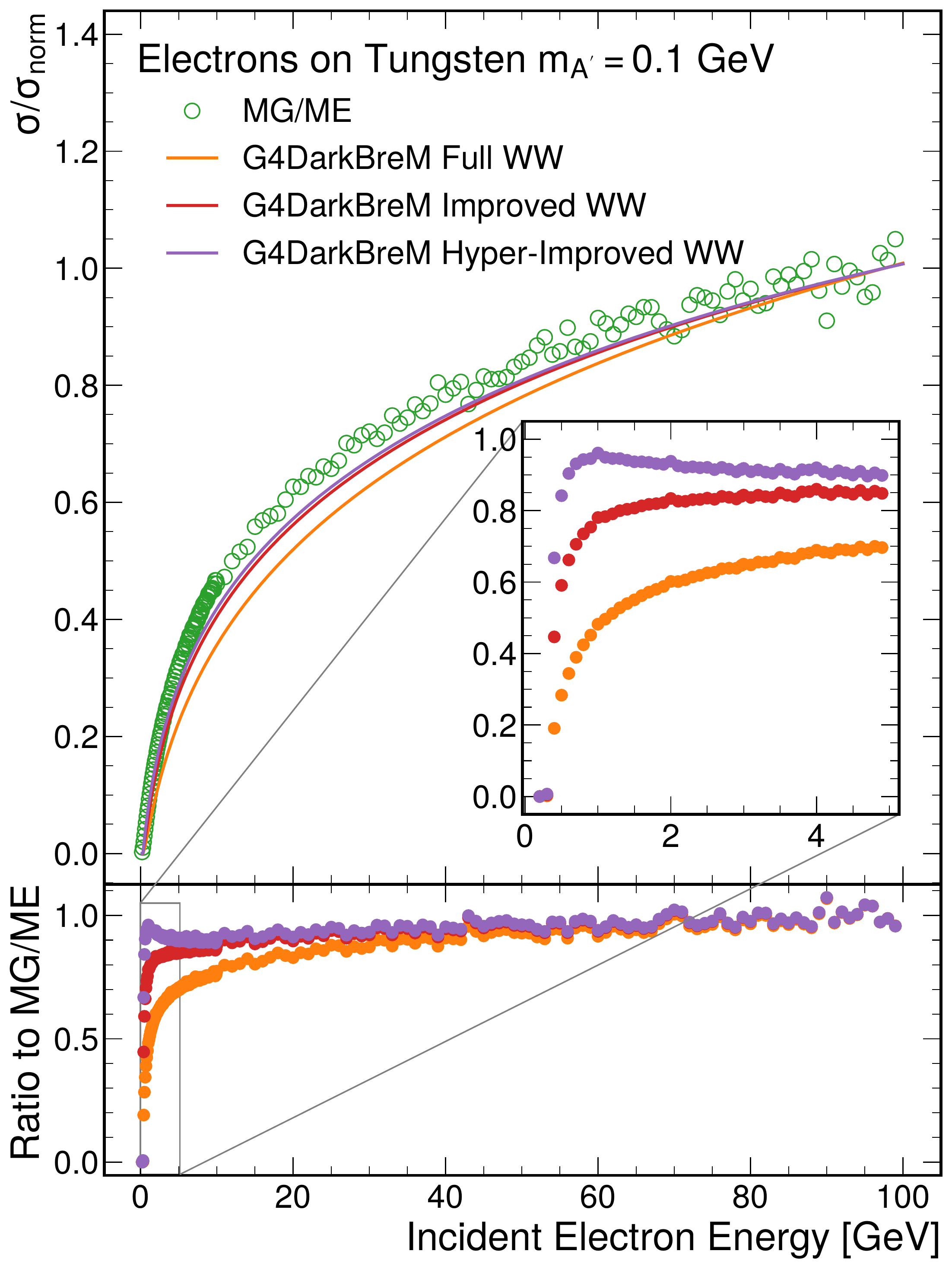}
    \end{subfigure}
    \begin{subfigure}[b]{0.45\textwidth}
        \centering
        \includegraphics[width=\textwidth]{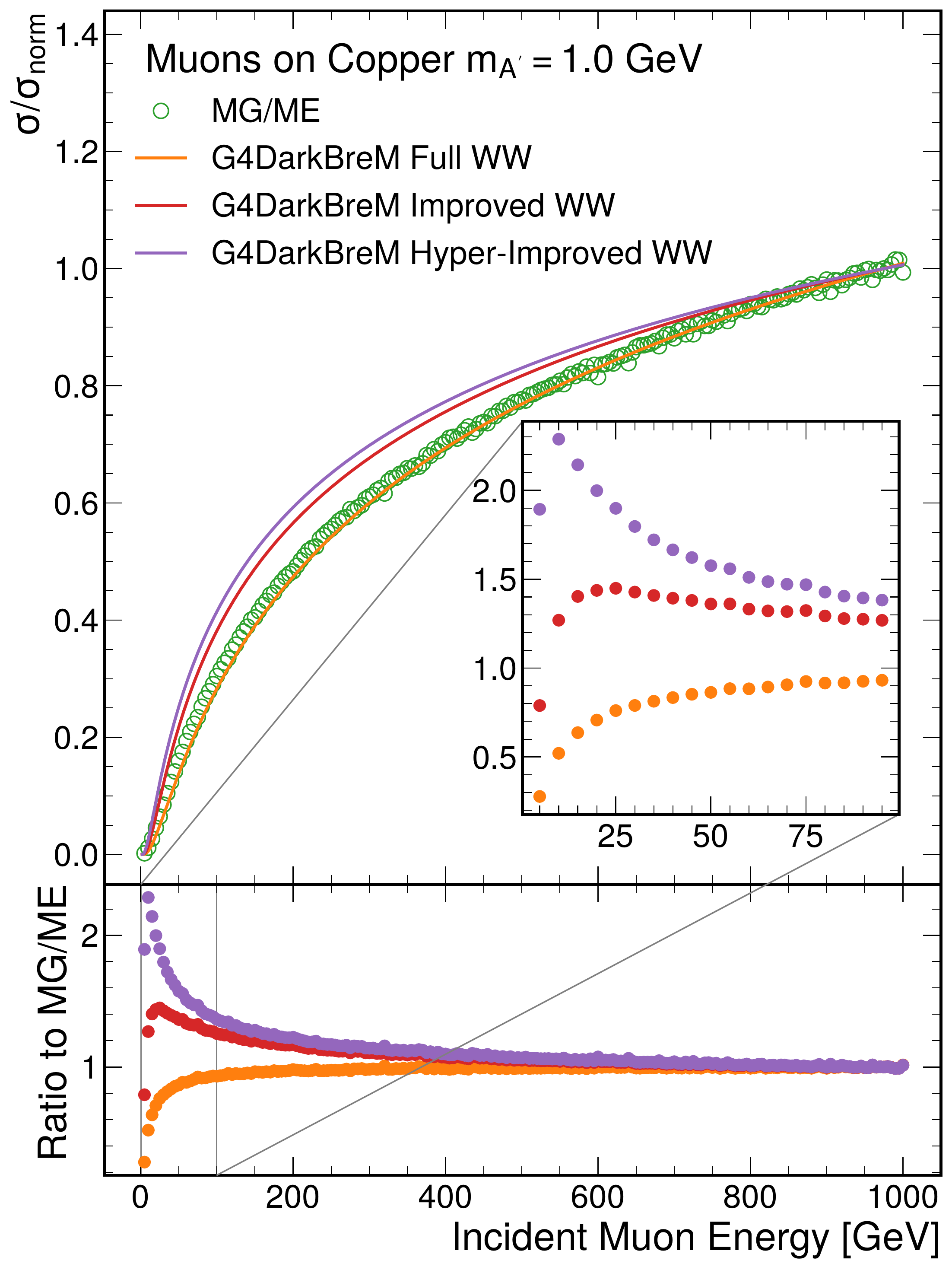}
    \end{subfigure}
    \caption{Comparison of the total \ac{db} cross-section as a function of incident lepton energy between the \ac{ww} approximation as calculated by the \acs{g4db} package (this work) and that calculated by \ac{mg} for electrons incident on tungsten with the mass of \aprime ($m_{\aprime}$) taken as $0.1$~GeV and muons incident on copper with $m_{\aprime}$ taken as $0.2$~GeV. For these two specific situations, the Improved \ac{ww} variation works best for electrons while the Full \ac{ww} works best for muons. The normalization factor $\sigma_{norm}$ is defined as the mean of the cross section samples for high energies ($E > 95$~GeV for electrons and $E > 950$~GeV for muons).}
    \label{fig:xsec_comp_appendix}
\end{figure}

Multiple variations of the \ac{ww} approximation exist, each best suited in a particular incident energy and lepton flavor domain. A subset of \ac{ww} approximations have been implemented in this package. 
None of these approximations are fully accurate and one should consider the incident energy and lepton flavor when choosing a suitable \ac{ww} approximation.
With $\epsilon$ being a free parameter and re-weighting events to remove separations in overall scale a simple procedure, the key goal of any of the total cross section approximations is to obtain a flat (or nearly flat) ratio with \ac{mg} within the energy region of interest.
Fig. \ref{fig:xsec_comp_appendix} shows a comparison of the three variations for the two situations studied in this work where we remove the absolute scale for easier comparison of the shape.
Other researchers \cite{Liu:2016mqv} have also seen an evolution of the \ac{ww} variations such that the one best suited to a particular situation depends on the beam energy, lepton mass, and dark photon mass.

In what follows, the flux factor $\chi(x,\theta)$ is calculated as
$$    
\chi = \int^{E_0^2}_{t_{min}} dt \left( \frac{Z^2a^4t^2}{(1+a^2t)^2(1+t/d)^2}+\frac{Za_p^4t^2}{(1+a_p^2t)^2(1+t/0.71)^8}\left(1+\frac{t(\mu_p^2-1)}{4m_p^2}\right)^2\right)\frac{t-t_{min}}{t^2}
$$
where
$$
a = \frac{111.0}{m_e Z^{1/3}}
\quad
a_p = \frac{773.0}{m_e Z^{2/3}}
\quad
d = \frac{0.164}{A^{2/3}}
$$
$$
\tilde{u} = -xE_0^2\theta^2 - m_A^2\frac{1-x}{x} - m_\ell^2x
\quad
t_{min} = \left(\frac{\tilde{u}}{2E_0(1-x)}\right)^2
$$
$E_0$ is the incoming lepton energy in GeV, $m_\ell$ is the mass of the electron in GeV, $m_A$ is the mass of the dark photon in GeV, $m_p = 0.938$ is the mass of the proton in GeV, $\mu_p = 2.79$ is the proton $\mu$, $A$ is the atomic mass of the target nucleus in amu, and $Z$ is the atomic number of the target nucleus.

\subsubsection{Full}
The \textit{Full \ac{ww}} approximation is motivated by the relatively large mass of the muon compared to the electron. While suffering from significant computation time by requiring a three-dimensional numerical integration, it is more accurate to the \ac{mg} total cross section when $\frac{m_A}{m_\ell}\lesssim 50$ compared to alternate \ac{ww} approximation schemes considered.

$$
\sigma = \frac{pb}{GeV} \int_0^{0.3} \int_0^{\min(1-m_\mu/E_0,1-m_A/E_0)} \frac{d\sigma}{dxd\theta}~dx~d\theta
$$

where 

$$    
\frac{d\sigma}{dx~d\cos\theta} = 2 \alpha_{EW}^3\epsilon^2 \sqrt{x^2E_0^2 - m_A^2}E_0(1-x) 
    \frac{\chi(x,\theta)}{\tilde{u}^2} \mathcal{A}^2
$$
$$
\mathcal{A}^2 = 2\frac{2-2x+x^2}{1-x}+\frac{4(m_A^2+2m_\mu^2)}{\tilde{u}^2}(\tilde{u}x + m_A^2(1-x) + m_\mu^2x^2)
$$
$\alpha_{EW} = 1/137$ is the fine-structure constant, $\epsilon$ is the dark photon mixing strength with the \ac{sm} photon, $pb/GeV = 3.894\times10^8$ is a conversion factor from GeV to pico-barns, and the other symbols are the same as the definition of $\chi(x,\theta)$ above.

\subsubsection{Improved}
The \textit{Improved \ac{ww}} approximation aims to simplify the numerical integral to two-dimensions by approximating the flux factor dependence on $\theta$ as the value at $\theta=0$, with a factor of two for the integral over $\cos\theta$. This leads to a gain in computation speed while maintaining reasonable accuracy. This approximation tends to over-estimate the absolute cross-section somewhat, but produces results which match better with MG/ME at $\frac{m_A}{m_\ell}\gtrsim 50$ compared with the Full \ac{ww}.  The approximation is also approximately 200 times faster in CPU-time than the Full \ac{ww}.

$$
\sigma = \frac{pb}{GeV} \int_0^{\min(1-m_\ell/E_0,1-m_A/E_0)} \chi(x,\theta=0)\frac{d\sigma}{dx}(x)dx
$$
where
$$
\frac{d\sigma}{dx}(x) = 4 \alpha_{EW}^3\epsilon^2 \sqrt{1-\frac{m_A^2}{E_0^2}}\frac{1-x+x^2/3}{m_A^2(1-x)/x+m_\ell^2x}
$$

\subsubsection{Hyper-Improved}
The \textit{Hyper-Improved \ac{ww}} approximation further simplifies the computation by treating the two-dimensional integral as two one-dimension integrals. The effective photon flux $\chi$ is a maxima at $x=1$ and $\theta=0$, which can be used over the entire phase space. 
$$
\sigma = \frac{pb}{GeV} \chi(x=1,\theta=0) \int_0^{\min(1-m_e/E_0,1-m_A/E_0)} \frac{d\sigma}{dx}(x)dx
$$
where the differential cross section $d\sigma/dx$ is the same as the Improved \ac{ww} above.

The Hyper-Improved algorithm leads to a significant gain in computation speed (factor of three relative to the Improved \ac{ww}) but at the cost of accuracy for lower lepton beam energies.  The approximation should not be used for $\frac{m_A}{m_\ell}< 50$, as the disagreement with MG/ME becomes very substantial for lower lepton energies.

\subsection{Scaling Method}
\label{sec:docs:method}

While energy fraction and transverse momentum are found to be accurate scaling variables, they do not define the sign of the outgoing longitudinal momentum. The simulation includes three different prescriptions (``scaling methods'') for resolving this ambiguity.  The default ``Forward Only'' method, used throughout this paper, always assumes that the lepton goes forward (i.e.~positive momentum along the beam axis). To solve this, the simulation always assumes that the lepton goes forwards and this scaling method is referred to as ``Forward Only". To recover these backwards scattering events, an alternate scaling method has been developed using the center of mass of the lepton/dark photon system. Study of \ac{mg} events with varying incident lepton energies showed that the transverse momentum of the center of mass of the scattered lepton and dark photon changes very slowly, and that its longitudinal momentum scaled linearly with incident energy. Applying this information, the ``CM Scaling" method works by finding the center of mass of the lepton and dark photon from the sampled \ac{mg} event, then reducing its longitudinal momentum and total energy by the difference between the sampled and desired incident energies. The sampled lepton is then boosted from the initial center of mass to the new one, and its outgoing angle in the new frame is collected. While using this angle preserves backwards-going events, the total energy was less reliable than using the Forward Only method. To reduce this, after finding the outgoing angle of the scattered lepton the total energy is sampled by preserving the fraction of kinetic energy in the same way as the forward only method. Even with this modification, the cost of keeping the backwards-going leptons is seen in slightly less accurate longitudinal momenta. The final method ``Undefined" is simply there to aid in studying the other scaling methods since it does not change the outgoing kinematics at all.

All of these methods have their advantages and dis-advantages, so they are all left in as options. Another study focusing on the detailed benefits of these different methods is required before making a firm conclusion about what should be done. In all cases, using a fine-grained incident lepton energy library ($< 10\%$ difference between two adjacent energy sample points) is \emph{heavily suggested} in order to minimize the discrepancies arising from any of these scaling methods.

\end{appendices}
\end{document}